\documentclass{aastex62}

\usepackage{subfigure}

\newcommand{\HI}{H\,\textsc{i}}
\newcommand{\DI}{D\,\textsc{i}}
\newcommand{\MgII}{\ion{Mg}{2}}

\newcommand{\Lya}{Ly$\alpha$}

\newcommand{\kms}{km s$^{-1}$}

\begin{document}

\title{Intrinsic \Lya~Profiles of High-Velocity G, K, and M Dwarfs  }

\author[0000-0002-1176-3391]{Allison Youngblood}
\affiliation{Exoplanets and Stellar Astrophysics Lab, NASA Goddard Space Flight Center, Greenbelt, MD 20771, USA}
\affiliation{Laboratory for Atmospheric and Space Physics, University of Colorado, 600 UCB, Boulder, CO 80309, USA}
\email{allison.a.youngblood@nasa.gov}

 \author[0000-0002-4489-0135]{J. Sebastian Pineda}
 \affiliation{Laboratory for Atmospheric and Space Physics, University of Colorado, 600 UCB, Boulder, CO 80309, USA}

 \author{Thomas Ayres}
 \affiliation{Center for Astrophysics and Space Astronomy, University of Colorado, 389 UCB, Boulder, CO 80309, USA}

 \author[0000-0002-1002-3674]{Kevin France}
 \affiliation{Laboratory for Atmospheric and Space Physics, University of Colorado, 600 UCB, Boulder, CO 80309, USA}
 \affiliation{Department of Astrophysical and Planetary Sciences, University of Colorado, UCB 389, Boulder, CO 80309, USA}
 \affiliation{Center for Astrophysics and Space Astronomy, University of Colorado, 389 UCB, Boulder, CO 80309, USA}

 \author{Jeffrey L. Linsky}
 \affiliation{JILA, University of Colorado and NIST, 440 UCB, Boulder, CO 80309, USA}

 \author[0000-0002-4998-0893]{Brian E. Wood}
 \affiliation{Naval Research Laboratory, Space Science Division, Washington, DC 20375, USA}

 \author[0000-0003-3786-3486]{Seth Redfield}
 \affiliation{Astronomy Department and Van Vleck Observatory, Wesleyan University, Middletown, CT 06459-0123, USA}

 \author[0000-0001-5347-7062]{Joshua E. Schlieder}
 \affiliation{Exoplanets and Stellar Astrophysics Lab, NASA Goddard Space Flight Center, Greenbelt, MD 20771, USA}

\begin{abstract}
Observations of \HI\ Lyman $\alpha$, the brightest UV emission line of late-type stars, are critical for understanding stellar chromospheres and transition regions, modeling photochemistry in exoplanet atmospheres, and measuring the abundances of neutral hydrogen and deuterium in the interstellar medium. Yet, \Lya\ observations are notoriously challenging due to severe attenuation from interstellar gas, hindering our understanding of this important emission line's basic morphology. We present high-resolution far- and near-UV spectroscopy of five G, K, and M dwarfs with radial velocities large enough to Doppler shift the stellar \Lya\ emission line away from much of the interstellar attenuation, allowing the line core to be directly observed. We detect self-reversal in the \Lya\ emission line core for all targets, and we show that the self-reversal depth decreases with increasing surface gravity. \MgII\ self-reversed emission line profiles provide some useful information to constrain the \Lya\ line core, but the differences are significant enough that \MgII\ cannot be used directly as an intrinsic \Lya\ template during reconstructions. We show that reconstructions that neglect self-reversal could overestimate intrinsic \Lya\ fluxes by as much as 60\%--100\% for G and K dwarfs and 40\%--170\% for M dwarfs. The five stars of our sample have low magnetic activity and sub-solar metallicity; a larger sample size is needed to determine how sensitive these results are to these factors.

\end{abstract}
\section{Introduction} \label{sec:Introduction}

\HI~Lyman $\alpha$ (\Lya; 1215.67 \AA) is the brightest UV emission line of late-type stars (F--M spectral types) and serves as an important diagnostic for stellar activity, the interstellar medium (ISM), and exoplanet atmospheres. However, bright geocoronal \Lya~emission (airglow) and resonant scattering from interstellar H I make direct observations of this intrinsically broad line's core impossible for the vast majority of stars. Measuring an accurate \Lya\ flux requires minimizing airglow through a narrow entrance aperture or slit and reconstructing the true flux from the observed line wings (e.g., \citealt{Wood2005}), or simply observing a star with large enough radial velocity to Doppler shift the entire line away from the ISM absorption \citep{Guinan2016,Schneider2019}. Unfortunately, most stars in the solar neighborhood have low radial velocities, ensuring that much of the stellar \Lya\ emission line coincides spectrally with the optically-thick ISM absorbers, and reconstructions are generally relied upon to obtain accurate intrinsic \Lya\ fluxes.

\Lya\ reconstructions often implement parametric models of the interstellar \HI~and \DI~absorption and the intrinsic stellar line profile (e.g., \citealt{Youngblood2016}), assuming a core line shape that is poorly constrained by observations and theory \citep{Bourrier2017_Kepler444,Fontenla2016,Peacock2019_trappist,Peacock_2019b,Tilipman2021}, although see \cite{Wood2005} and \cite{Zhang2021} for non-parametric examples. Our lack of knowledge of the true \Lya\ line core shape limits the accuracy of our reconstructed fluxes and therefore our exoplanet atmosphere models, for which \Lya\ flux and its flux density profile are critical inputs. \Lya~controls the energy balance of \HI\ atoms in planetary atmospheres; these atoms are generally cold and absorb within $\pm$20 \kms\ of line center \citep{Emerich2005,Tian2009_coronalholes}. \Lya\ also ionizes NO and photodissociates important molecules such as H$_2$O, CO$_2$, and CH$_4$, thereby influencing the chemical balance of O$_2$ and O$_3$ in planetary atmospheres.

\begin{figure}
         \centering
          \includegraphics[width=0.6\textwidth]{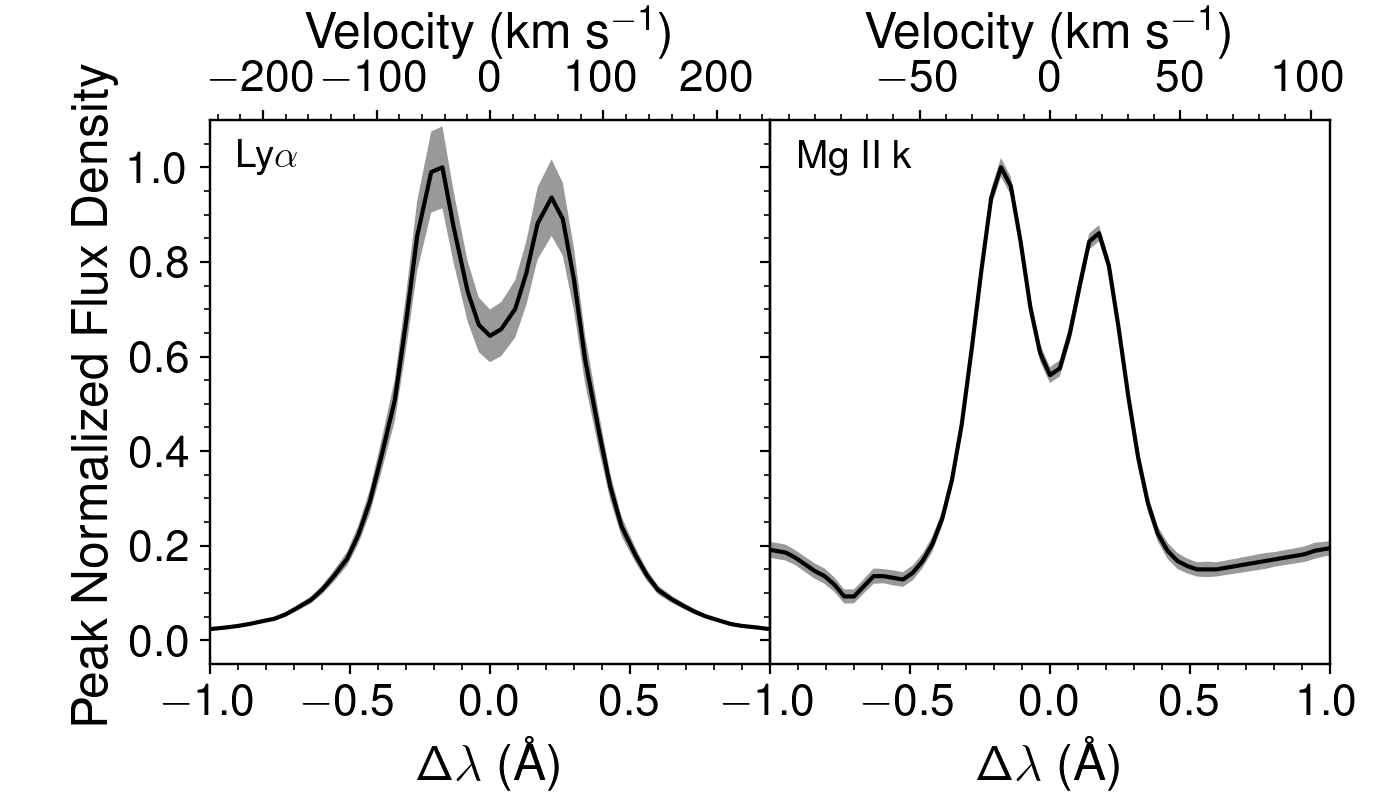}
          \caption{Left: The average quiet-Sun \Lya\ profile derived from \textit{SOHO}/SUMER data ($R\sim$14,000; \citealt{Wilhelm1997}) taken on 2008 June 24-26 \citep{Gunar2020}. Right: The average quiet-Sun Mg II k solar profile derived from \textit{IRIS} data ($R\sim$35,000; \citealt{DePontieu2014}) taken between 2019 April and 2020 September \citep{Gunar2021}. Both profiles are peak normalized, and the grey shading represents the reported 1-$\sigma$ uncertainties. See \cite{Fontenla1988} and \cite{Schmit2015} for examples of solar \Lya\ and \MgII\ profiles over magnetically-active regions such as plage and sunspots; these profiles generally exhibit shallower self-reversal than the quiet Sun.}
          \label{fig:solar_LyA_MgII}
  \end{figure}

The Sun is the only star whose entire \Lya\ profile has been observed at high spectral resolution and high signal-to-noise without obscuring effects from the ISM or airglow. The quiet Sun's average \Lya\ profile is Voigt-like except in the line core where it exhibits self-reversal, with a slight asymmetry making the blue peak brighter than the red peak (Figure~\ref{fig:solar_LyA_MgII}; \citealt{Fontenla1988,Gunar2020}). The self-reversal is shallower or even disappears completely over magnetically active regions such as plage or sunspots \citep{Fontenla1988}.

Self-reversed emission lines display broad intensity dips in their cores and are characteristic of optically-thick chromospheric scattering lines such as Mg II, Ca II, Na I, and H-alpha \citep{Linsky1979a,Short1998}. The outermost layers of the chromosphere where \Lya\ forms have low density and large mean free paths against collisions and thus strongly depart from local thermodynamic equilibrium (LTE). Therefore, despite the increasing temperature with altitude, the emissivity at high altitudes is depressed by the low efficiency of thermal emission and photon scattering losses, mainly to the line wings. The density increases in the mid-chromosphere, reducing scattering lengths and increasing the efficiency of thermal emission due to increasing collisional excitation and quenching. In the low chromosphere, where the density is even higher, collisional processes dominate and the emissivity has fully thermalized.

Roughly, the emergent intensity profile maps to the emissivity corresponding to optical depth unity; further away from line center the $\tau$=1 surface moves deeper in the atmosphere. Near line center, the intensity is reduced by the depressed high-altitude emissivity, but then slightly away from line center, the intensity rises as the emissivity couples better to the local thermal collisional excitation at mid-altitudes. The intensity declines again in the outer line core, reflecting the decreasing temperatures, and thus also reduced thermal emission, at lower altitudes in the chromosphere. If a chromosphere has higher density, as in a solar magnetic plage region or a high surface gravity M dwarf, the collisional coupling in the higher layers is enhanced, so that the central reversal exhibits lower contrast, or disappears completely in extreme cases (see \citealt{Ayres1979}).

Stellar observations of various chromospheric emission lines, including \Lya, indicate that \Lya\ self-reversal deepens with increasing stellar mass. For example, high-resolution spectra of optically-thick chromospheric emission lines that are less affected by ISM attenuation than \Lya, such as \MgII\ and Ca II, show that solar type stars clearly have deep self-reversal \citep{Linsky1979a} while M dwarfs have little to no self-reversal \citep{Wood2005}. Medium resolution \Lya\ spectra of high radial velocity K and M dwarfs \citep{Guinan2016,Youngblood2016,Bourrier2017_Kepler444,Schneider2019} and other stars without a directly-observable line core (e.g., \citealt{Wood2005,GarciaMunoz2020,Carleo2021,Bourrier2017_Kepler444}) generally support the self-reversal trend with mass. This suggests stronger collisional coupling with the chromospheric temperature rise due to larger chromospheric densities in M dwarfs, which would be a natural outcome of their higher surface gravities compared with warmer, lower gravity G dwarfs like the Sun. Also, many M dwarfs display higher magnetic activity levels than typical G dwarfs, partially due to their longer lives and slower rotational evolution (e.g., \citealt{Reiners2012}). Higher activity in the well-vetted solar example (e.g., plage regions versus the quiet Sun) is associated with higher chromospheric densities, which presumably also would be the case for active stars. 

Nevertheless, stellar models currently struggle to replicate the shallow self-reversals expected for M dwarfs, instead producing very deep reversals
\citep{Fontenla2016,Peacock_2019b,Peacock2019_trappist,Tilipman2021}. \cite{Peacock_2019b} note that missing physics from their 1D models like a corona and ambipolar diffusion could be creating inaccuracies in the hydrogen ionization, and \cite{Judge2020} point out the importance of 3D effects in controlling the appearance of the \Lya\ line.

We have observed five new high radial velocity stars ranging from spectral type G-M at high resolution in order to spectrally resolve and quantitatively measure \Lya\ self-reversal and uncover trends with spectral type. We have obtained contemporaneous high-resolution \MgII\ spectroscopy to ascertain if that line's core serves as a suitable proxy for \Lya's. Section~\ref{sec:ObservationsReductions} describes the target selection, observations and data reduction, Section~\ref{sec:linefitting} details the reconstruction and fitting technique, and Section~\ref{sec:Results} presents the results. Section~\ref{sec:Discussion} discusses the implications of the results, and Section~\ref{sec:Conclusions} concludes.

\section{Observations \& Reductions} \label{sec:ObservationsReductions}

Observations were obtained with the \emph{Hubble Space Telescope} (\emph{HST}) Space Telescope Imaging Spectrograph (STIS) as part of HST-GO-15190 between 2018 August and 2020 January. Table~\ref{Tab:Observations} lists our targets, their spectral types, distances, radial velocities, surface gravities, and details about the instrument setup and observing time. The five targets were selected based on the following criteria: radial velocity $|V_{\rm radial}|$ $>$ 80 \kms, $d\leq$6 pc, and spectral types G-M. The radial velocity criterion ensures that the core of the emission line will be Doppler-shifted away from the interstellar \ion{H}{1} attenuation trough, the distance criterion ensures sufficient signal-to-noise (S/N) at high spectral resolution with STIS, and the spectral type range allows us to explore the shape of the line cores as a function of stellar mass. 

The target list includes one G dwarf (82 Eri), one K dwarf (HD 191408), and three M dwarfs (Barnard's Star, Kapteyn's Star, and GJ 411). Here we provide details for each star, including the determination of the stellar surface gravity ($GM$/$R^{2}$) that will be used in our analysis. Measuring masses for individual stars is notoriously challenging and uncertain, and we elected to adopt masses derived from mass-luminosity relations. Radii are from direct interferometric measurements when available and radius-luminosity relations otherwise. As described in the next paragraphs, our sample size is small and skewed toward stars with low activity and low metallicity. The impacts of this important limitation on our results is discussed in Section~\ref{sec:Discussion}.  

\paragraph{82 Eri} 

This G8 V star (6.04 pc, +87.9 \kms) with four confirmed exoplanets \citep{Pepe2011,Feng2017} is likely field age ($\sim$5 Gyr) based on its kinematics \citep{Gagne2018}. \cite{Ghezzi2010} estimate via isochrones that the star is $>$13 Gyr old and determine its metallicity [Fe/H] = -0.42$\pm$0.02. To determine 82 Eri's surface gravity, we adopt the mass-luminosity and radius-luminosity relations of \cite{Eker2018}, which are valid for 0.72-1.05 M$_{\odot}$.  The stellar bolometric luminosity was derived from the V band magnitude \citep{Ducati2002}, the distance from Table~\ref{Tab:Observations}, and a bolometric correction from \cite{Pecaut2013}\footnote{http://www.pas.rochester.edu/$\sim$emamajek/EEM\_dwarf\_UBVIJHK\_colors\_Teff.txt}. There are significant differences between the mass and radius we derive (0.928$\pm$0.007 M$_{\odot}$ and 0.897$\pm$0.176 R$_{\odot}$) and those from \cite{Ghezzi2010}, but our log $g$ value (4.50$\pm$0.08) overlaps with that work's value at the 1-$\sigma$ level. Note that 82 Eri's metallicity falls outside the calibration range of the \cite{Eker2018} scaling relations, meaning that the stellar surface gravity may be affected by systematic uncertainty.

\paragraph{HD 191408}

This K2.5V star (6.01 pc, -129.3 \kms) is also likely field age based on its kinematics \citep{Gagne2018}. \cite{Ghezzi2010} estimate its age via isochrones as 9-14 Gyr and measure [Fe/H] = -0.56$\pm$0.04. We follow the approach outlined for 82 Eri to determine HD 191408's surface gravity using the \cite{Eker2018} relations, despite HD 191408's low metallicity. We used the V band magnitude from \cite{Zacharias2013} to determine the bolometric luminosity. Like for 82 Eri, there are significant differences between the mass and radius we derive (0.805$\pm$0.016 M$_{\odot}$, 0.744$\pm$0.176 R$_{\odot}$) and those from \cite{Ghezzi2010}, but our log $g$ value (4.60$\pm$0.10) overlaps with that work's at the 1-$\sigma$ level.

\paragraph{Kapteyn's Star}

This M1 subdwarf (3.93 pc, +245 \kms) with one confirmed exoplanet \citep{Anglada-Escude2014} is a halo star possibly born in the $\omega$ Cen globular cluster \citep{Kotoneva2005}, which has an average metallicity [Fe/H] = -1.35 and is approximately 11.5 Gyr old \citep{MarinFranch2009,Forbes2010}. Kapteyn's Star's metallicity has been measured as [Fe/H] = -0.88$\pm$0.08 \citep{Neves2013}. We combine the $K_{\rm S}$ band magnitude from 2MASS \citep{Cutri2003} and the distance reported in Table~\ref{Tab:Observations} with the \cite{Kesseli2019} M dwarf radius-luminosity relation and the \cite{Mann2019} M dwarf mass-luminosity relations. The \cite{Kesseli2019} radius relation is valid for metallicities [Fe/H] $>$ -2.0 dex, and the \cite{Mann2019} relation is only explicitly valid for [Fe/H] $>$ -0.6. However, \cite{Mann2019} estimate that their masses are insensitive to metallicity (0.0\%$\pm$2.2\% per dex), and therefore the relation is likely appropriate for Kapteyn's Star. We find M=0.279$\pm$0.008 M$_{\odot}$, R=0.283$\pm$0.094 R$_{\odot}$, and combine them to derive log $g$ = 4.98$\pm$0.13.

\paragraph{GJ 411}

This M2V star (2.55 pc, -84.7 \kms) with two confirmed exoplanets \citep{Diaz2019,Rosenthal2021} is likely field age ($\sim$5 Gyr) based on its kinematics\footnote{http://www.exoplanetes.umontreal.ca/banyan/} \citep{Gagne2018} and has metallicity [Fe/H] = -0.41$\pm$0.17 \citep{RojasAyala2012}. We use the mass and radius posterior distributions (0.389$\pm$0.008 M$_{\odot}$ and 0.392$\pm$0.004 R$_{\odot}$) from \cite{Pineda2021_MUSS} to derive log $g$ = 4.84$\pm$0.01. \cite{Pineda2021_MUSS} used the angular diameter from \cite{Boyajian2012} and a new empirical mass-radius relationship to derive these quantities. 

\paragraph{Barnard's Star}

This M4V star (1.83 pc, -110.4 \kms) is approximately 7-10 Gyr old \citep{Ribas2018} with metallicity [Fe/H] = -0.52$\pm$0.08 \citep{Neves2013}. We use the mass and radius posteriors (0.161$\pm$0.004 M$_{\odot}$ and 0.187$\pm$0.001 R$_{\odot}$) from \cite{Pineda2021_MUSS} to derive log $g$ = 5.10$\pm$0.01. We note that we tested the scaling relations used for Kapteyn's Star on Barnard's Star and GJ 411 and find masses and radii consistent with those adopted in this work. \\

In order to spectrally resolve any self-reversal in the line cores of the \Lya\ (1216 \AA) and \MgII\ (2796, 2802 \AA) chromospheric emission lines, we selected the STIS E140M and E230H gratings, respectively. \Lya\ is an intrinsically broad emission line in the spectra of late type stars, with FWHM of order 100 \kms, while \MgII\ is intrinsically narrow with FWHM of order 10 \kms. The exact intrinsic widths of these chromospheric emission lines positively correlate with stellar luminosity (i.e., the Wilson-Bappu effect; \citealt{Wilson1957}). The E140M observations were centered at 1425 \AA\ and cover 1150-1700 \AA\ at spectral resolving power $R = \lambda$/$\Delta \lambda \approx$ 45,800, which corresponds to a resolution element of 0.027 \AA\ or 6.6 km s$^{-1}$ at \Lya. The E230H observations were centered at 2713 \AA\ and cover 2574-2851 \AA\ at $R \approx$ 100,000, which corresponds to a resolution element of 0.028 \AA\ or 3 km s$^{-1}$ at \MgII.

Each target was observed in a single visit in order to mitigate the impact of stellar variability on the \Lya\ to \MgII\ comparisons, and all exposures were obtained in time-tag mode. No flares were observed. The guide star acquisition of HD 95735 failed on 2019 May 15 and was retaken via HOPR 91550 on 2020 January 14 with a wider slit. We only present data here from the 2020 observation. 

The STIS E140M and E230H echellegrams were post-processed starting from the standard CALSTIS pipeline \texttt{x1d} files, which contain tabulations of wavelengths, fluxes, photometric errors, and data quality flags for each of the echelle orders. Several of the post-pipeline ASTRAL (Advanced Spectral Library Project\footnote{https://casa.colorado.edu/$\sim$ayres/ASTRAL/}) procedures were utilized to merge the overlapping wavelengths in each order of each echellegram, and combine spectra taken in the same visit with the same instrumental setup. Two of the targets, 82~Eri and HD~191408, had sufficiently bright continuum emission in the NUV region to make use of the ASTRAL blaze correction algorithm, which iteratively finds a blaze shift that optimally balances the fluxes in the overlap zones between echelle orders, averaged over the echellegram.  However, the rest of the NUV observations, and all the FUV, were too faint for this option to be utilized.

We determined that the effect of \Lya\ scattered light on the spectrum was minimal, due largely to the intrinsic faintness of the target stars. Airglow contamination was also minimal, because we restricted each target's visit to occur during the time of year when Earth's projected velocity toward our target sight lines coincides with the expected spectral width and velocity of the ISM absorption. The narrow \HI\ geocoronal emission lines appear in the ISM absorption troughs where there is zero signal from the star. A Gaussian was fitted to these airglow features and then subtracted.

\begin{deluxetable}{lcccccccc}
\tablecolumns{9}
\tablewidth{0pt}
\tablecaption{Targets and Observations \label{Tab:Observations}}
\tablehead{\colhead{Target} &
                  \colhead{Spectral} &
                  \colhead{$d$} &
                  \colhead{$V_{\rm radial}$} &
                  \colhead{log $g$} &
                  \colhead{Observation} &
                  \colhead{Grating} &
                  \colhead{Aperture} &
                  \colhead{Exp.} \\
                  \colhead{Name} &
                  \colhead{Type} &
                  \colhead{(pc)} &
                  \colhead{(km s$^{-1}$)} & 
                  \colhead{(cgs)} &
                  \colhead{Date} &
                  \colhead{} &
                  \colhead{} &
                  \colhead{Time (s)} 
                  }
\startdata
82 Eri &  G8V & 6.04 & +87.9 & 4.50$\pm$0.08 & 2018-Dec-15 & E140M & 0.2\arcsec$\times$0.06\arcsec\ & 3919 \\
     & & & & & 2018-Dec-15 & E230H & 0.2\arcsec$\times$0.2\arcsec\ &   503    \\
     \hline
HD 191408  & K2.5V & 6.01 & -129.3 & 4.60$\pm$0.10 & 2018-Sep-04 & E140M & 0.2\arcsec$\times$0.06\arcsec\ & 4064 \\
     & & & & & 2018-Sep-04 & E230H & 0.2\arcsec$\times$0.2\arcsec\ &   306    \\
     \hline
Kapteyn's  & sdM1 & 3.93 & +245.0 & 4.98$\pm$0.13 & 2019-Apr-03 & E140M & 0.2\arcsec$\times$0.06\arcsec\ & 6312 \\ 
Star     & & & & & 2019-Apr-03 & E230H & 0.2\arcsec$\times$0.2\arcsec\ &  976     \\
\hline
GJ 411  & M2V & 2.55 & -84.7 &  4.84$\pm$0.01 & 2020-Jan-14 & E140M & 0.2\arcsec$\times$0.2\arcsec\ & 7193 \\ 
     & & & & & 2020-Jan-14 &E230H & 0.2\arcsec$\times$0.2\arcsec\ &   316    \\
     \hline
Barnard's  & M4V & 1.83 & -110.4 & 5.10$\pm$0.01 & 2018-Aug-12 & E140M & 0.2\arcsec$\times$0.06\arcsec\ & 9195 \\
Star     & & & & & 2018-Aug-12 & E230H & 0.2\arcsec$\times$0.2\arcsec\ & 1029      \\
\enddata
\tablecomments{All distances ($d$) and radial velocities ($V_{\rm radial}$) are from \emph{Gaia} DR2 \citep{Gaia2016,GaiaDR2,Soubiran2018} except for GJ 411's $V_{\rm radial}$ \citep{Nidever2002}. The central wavelength setting for all E140M observations is 1425 \AA\ and for all E230H observations is 2713 \AA. Note that all observations and discussion of HD 191408 refer to the primary component of the binary. HD 191408 B is a K or M dwarf \citep{Bidelman1985,Howard2016} separated 4.06\arcsec\ from the primary according to \emph{Gaia}.}
\end{deluxetable}

  \begin{figure}
         \centering
          \includegraphics[width=\textwidth]{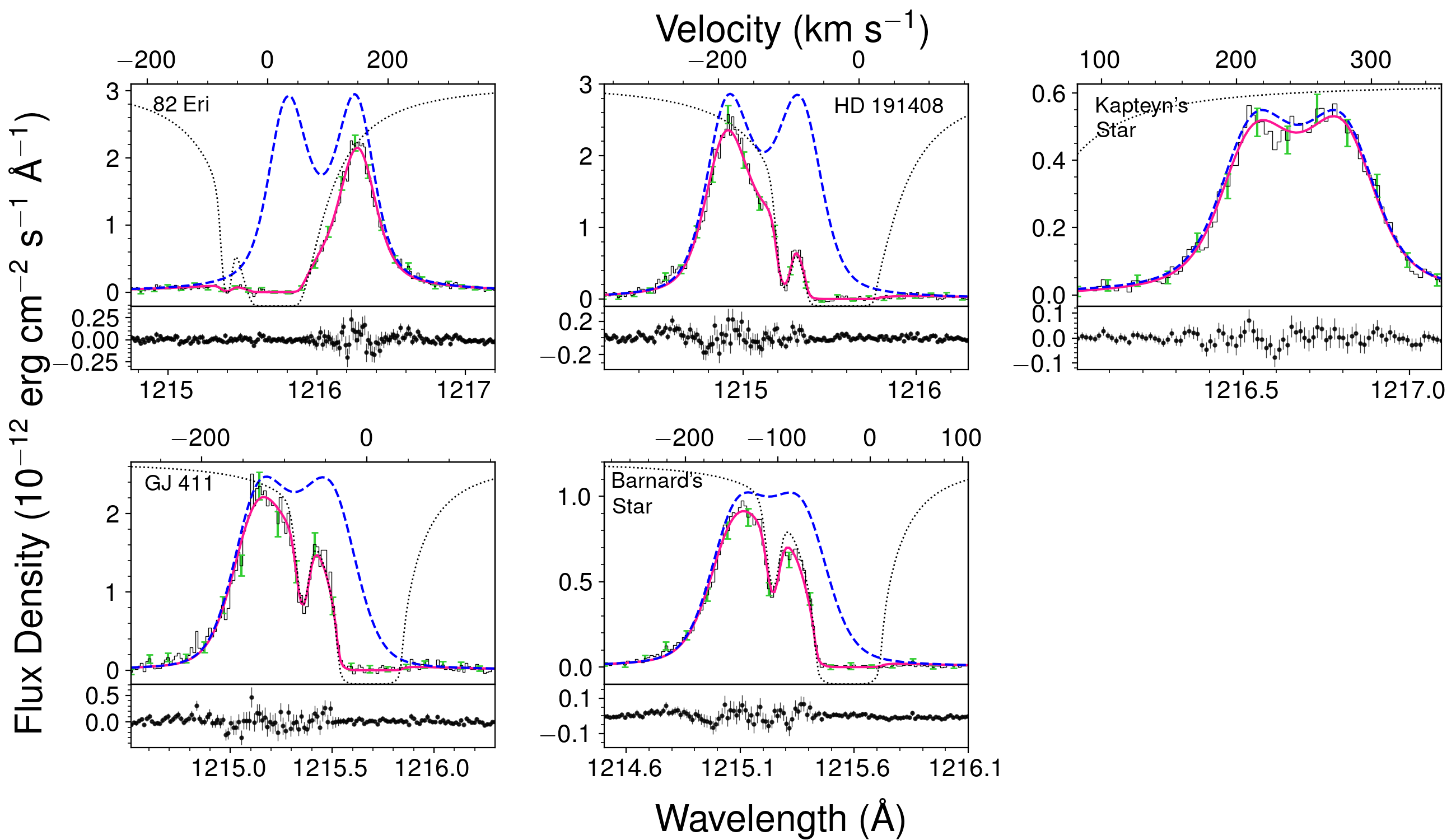}
          \caption{\Lya\ reconstructions for the target stars. The black histogram with green error bars show the STIS E140M data (post-airglow subtraction) with 1-$\sigma$ uncertainties shown every seventh data point for visual clarity. The dashed blue line shows the reconstructed intrinsic profile, which when multiplied by the dotted black line showing the ISM attenuation (with values ranging from 0-1 scaled to the vertical range of the plot) yields the median best fit profile (pink line). The small panels below each fit result show the residuals (data - model). Heliocentric velocities are shown for reference. The line cores of all five stars are clearly detected in the STIS spectra.}
          \label{fig:LyA_reconstruciton_results}
  \end{figure}
  
  \begin{figure}
         \centering
          \includegraphics[width=0.6\textwidth]{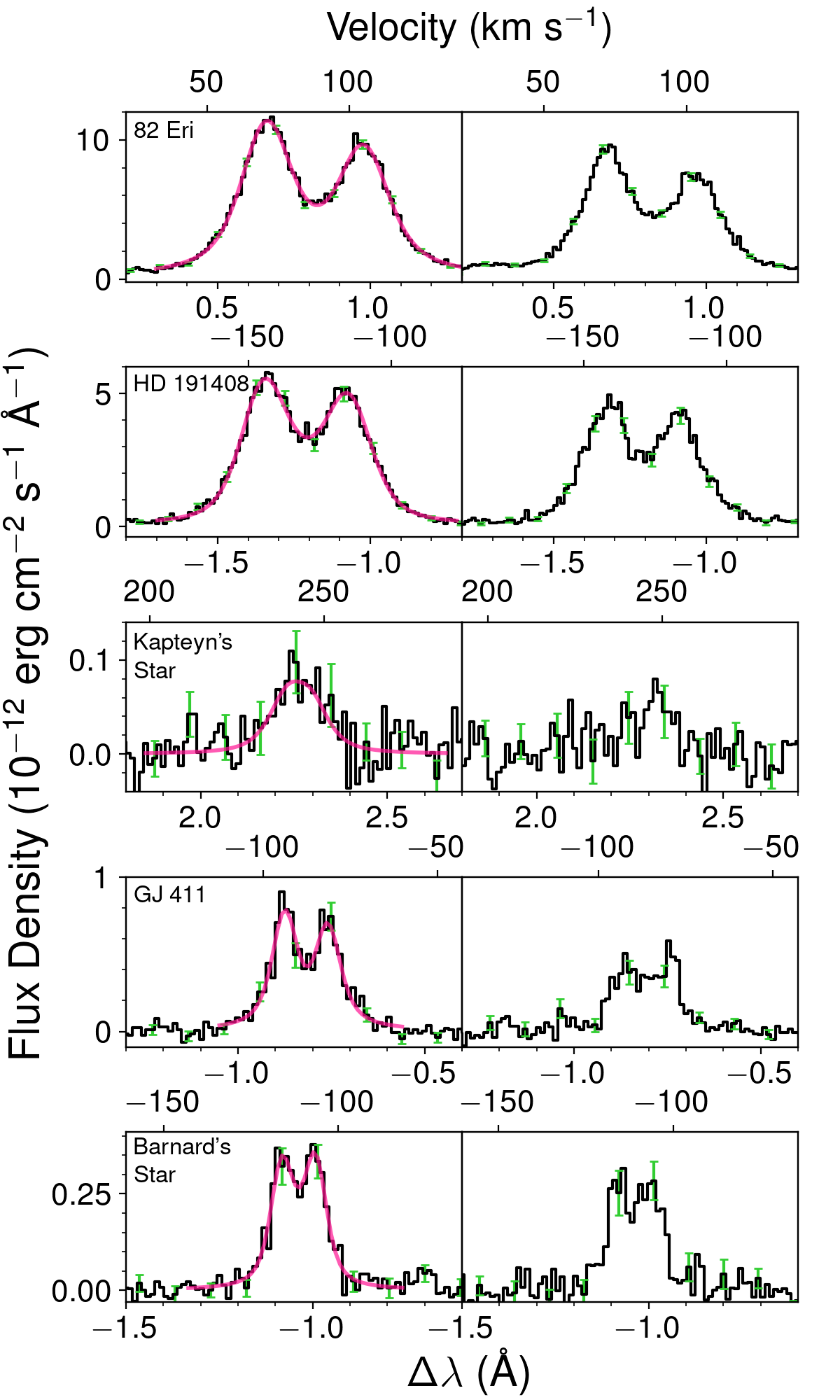}
          \caption{The Mg II k \& h profiles as observed for our target stars. The black histogram and green error bars show the STIS E230H data with 1-$\sigma$ uncertainties shown every eighth data point for visual clarity. The bottom horizontal axes show the wavelength arrays offset from the h (left panels) and k (right panels) transitions' rest wavelengths (2796.3543 \AA\ and 2803.5315 \AA, respectively) for visual clarity. The pink lines show the fits to the k line. Heliocentric velocities are shown for reference.
}
          \label{fig:MgII_data}
  \end{figure}

  \begin{figure}
         \centering
          \includegraphics[width=0.5\textwidth]{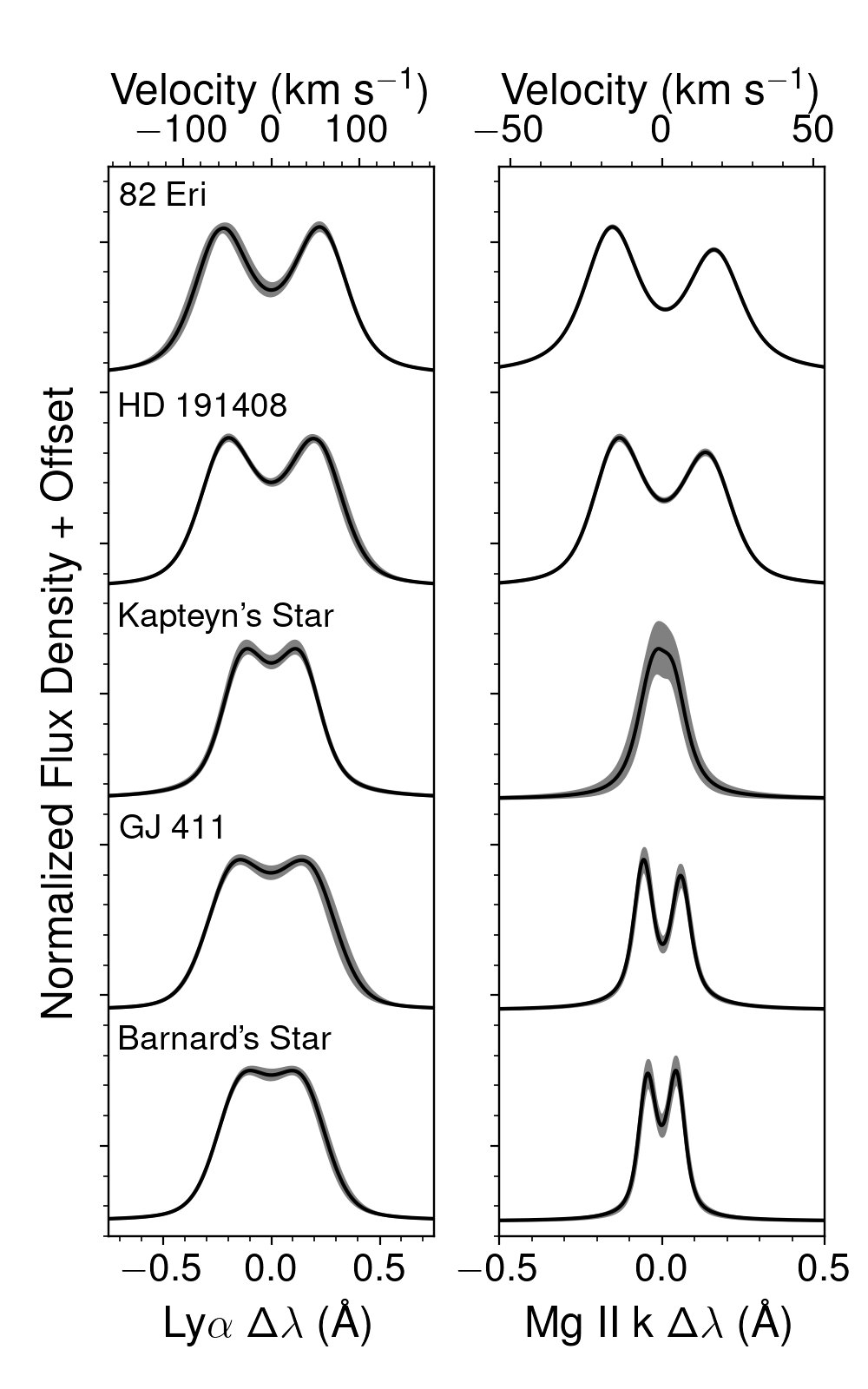}
          \caption{The un-convolved intrinsic \Lya\ (left) and \MgII\ k (right) profiles of the targets. The black lines show the median value (normalized by its peak value) as a function of wavelength from the ensemble MCMC samples and the gray shading shows the 68\% confidence interval. Velocities in the stellar reference frame are shown.}
          \label{fig:LyA_MgII_comparison}
  \end{figure}

\section{\Lya\ and \MgII\ analysis} \label{sec:linefitting}

\subsection{Description of model and fitting routines}

Interstellar hydrogen and deuterium absorption is hundreds of km s$^{-1}$ wide, even for nearby stars, meaning that all of our high-radial velocity targets require correction for ISM attenuation. Note that, unlike \Lya, the \MgII\ emission lines of our targets are not affected by ISM attenuation; the stellar radial velocities are sufficiently large to Doppler shift the ISM \MgII\ absorption away from the stellar emission line. This difference between \Lya\ and \MgII\ is due to the combined effect of the narrower intrinsic line width of \MgII\ and the lower abundance of Mg in the ISM. In order to recover the intrinsic \Lya\ profiles from our E140M spectra, we model the observed \Lya\ profiles as the product of two components: the stellar emission component and the ISM absorption component. Previous works have found that a Voigt profile fits the Gaussian-like core and Lorentzian-like wings of \Lya\ well \citep{GarciaMunoz2020,Carleo2021,Youngblood2021}, so we select a Voigt profile as the parameterization of the emission component without any self-absorption. We also tested and ruled out the \MgII\ line as a scaleable template for the entire \Lya\ emission line, including self-reversal. The \MgII\ templates were created by smoothing the observed k line spectra, then the templates were shifted in velocity space and their widths and amplitudes were scaled to match the data. No satisfactory fits were achieved; when scaling the \MgII\ templates to fit the width of the \Lya\ lines, the separations between the two peaks of the self-reversed profile became too wide. Note that this does not imply there is no correlation between the fluxes of the \Lya\ and \MgII\ emission lines, which has been clearly demonstrated in the literature (see \citealt{Wood2005}).

We tested several parameterizations for self-reversal of the Voigt emission line including a Gaussian or Voigt in absorption. We performed tests where the velocity centroid was either fixed to the emission line's centroid or free to vary and the self-reversal FWHM was independent of the emission line FWHM. The Gaussian and Voigt in absorption performed equally well according to the Bayesian Information Criterion (BIC), and we found that allowing the centroid to vary freely only improved the BIC for one star.  We also fit the line profiles with no self-absorption, and in all cases find that self-absorption is clearly warranted to substantially improve the fit quality.

We select the Voigt in absorption for the self-reversal because of its ability to fit the data well, and because \cite{Cowan1948} used a similar model to explain self-reversal observed in laboratory arc lamps. Their model requires the self-reversal to exactly mimic the reversal-free emission line shape and is given by:

\begin{equation} \label{eq:em}
    F_{emission}^{\lambda} = \mathcal{V}^{\lambda}(V_{\rm radial},A,FWHM_{\rm L},FWHM_{\rm G}) \cdot \exp \Big(-p \mathcal{V}^{\lambda}_{\rm norm} \Big),
\end{equation}

\noindent where $\mathcal{V}^{\lambda}$ is the Voigt profile (\texttt{astropy Voigt1D} function; \citealt{McLean1994}), $\mathcal{V}^{\lambda}_{\rm norm}$ is the peak-normalized $\mathcal{V}^{\lambda}$, and $p$ is the unitless self-absorption parameter. Here we refer to self-absorption as a general condition of which self-reversal is an extreme case: $p$=0 corresponds to the absence of any self-absorption (i.e., the line intensity shape of a chromosphere in local thermodynamic equilibrium), 0$<p<$1 yields slightly flat-topped emission lines, and the transition from self-absorption to self-reversal (a dip in intensity in the line core) occurs at $p$=1. Self-reversal deepens with increasing $p$ for values $>$1. The free parameters of $\mathcal{V}^{\lambda}$ are the radial velocity of the emission line, $V_{\rm radial}$ (\kms), the Lorentzian amplitude, $A$ (erg cm$^{-2}$ s$^{-1}$ \AA$^{-1}$), and the full width at half maximum values for the Lorentzian and Gaussian components, $FWHM_{\rm L}$, and $FWHM_{\rm G}$ (\kms).

We assume a single interstellar absorbing cloud. Most sight lines intersect multiple clouds \citep{Redfield2008}, but often multiple clouds can be approximated with a single absorption component \citep{Youngblood2016} by inflating the fitted column densities and Doppler broadening parameters.  The ISM component is modeled as a single Voigt profile in absorption for \ion{H}{1} and \ion{D}{1} each using the code \texttt{lyapy}\footnote{https://github.com/allisony/lyapy} \citep{Youngblood2016}:

\begin{equation} \label{eq:abs}
    F_{\rm absorption}^{\lambda} = \mathcal{V}^{\lambda}(V_{\rm HI}, N(HI), b_{HI}) \times \mathcal{V}^{\lambda}(V_{\rm DI}, N(DI), b_{DI}),    
\end{equation}

\noindent where $V_{\rm HI}$ is the radial velocity (\kms), assumed to be identical for both \ion{H}{1} (1215.67 \AA) and \ion{D}{1} (1215.34 \AA); $N(HI)$ is the hydrogen column density (cm$^{-2}$); $N$(\ion{D}{1}) is the deuterium column density, taken to be 1.5$\times$10$^{-5}$ $\times$ $N$(\ion{H}{1}) \citep{Linsky2006}; and $b$ is the Doppler broadening parameter. We link $b_{\rm HI}$ and $b_{\rm DI}$ under the assumption of thermal equilibrium so that $b_{\rm DI}$ = $b_{\rm HI}$/$\sqrt{2}$, but note that in reality there is a small turbulent broadening component. 

Absorption from the stellar astrosphere or the heliosphere can appear in the blue and red wings, respectively, of the ISM \HI\ absorption troughs (see \citealt{Wood2021} and references therein), affecting any \Lya\ reconstructions that do not explicitly model this extra absorption. \cite{Wood2021} searched for astrospheric absorption in the \Lya\ spectra of GJ 411 and Barnard's Star presented here, but found none. Heliospheric absorption is likely present in the red wing of the ISM \HI\ absorption trough of Barnard's Star, given that star's close proximity and position in the sky only 23$^{\circ}$ from the upwind direction of the ISM flow seen by the Sun \citep{Wood2005}. However, including heliospheric absorption in our model profile would not significantly impact the reconstruction given the low signal-to-noise of the spectrum near the expected absorption. For our other targets, excess absorption from an astrosphere or the heliosphere is not seen, although a rigorous analysis is beyond the scope of this paper.

To model the observed \Lya\ profile, we multiply the emission and absorption models (Equations~\ref{eq:em} and \ref{eq:abs}) and convolve with the instrument line spread function (LSF) provided by STScI\footnote{https://www.stsci.edu/hst/instrumentation/stis/performance/spectralresolution} for the appropriate grating and slit size:

\begin{equation} \label{eq:model}
    F^{\lambda} = (\mathcal{V}_{\rm emission} \times \mathcal{V}_{\rm absorption}) \circledast LSF.
\end{equation}

We fit Equation~\ref{eq:model} to the data using a likelihood-based Bayesian calculation and the \texttt{emcee} Markov Chain Monte Carlo implementation \citep{Foreman-Mackey2013}. \texttt{emcee} maximizes the sum of the logarithm of our parameters' prior probabilities and the logarithm of a likelihood function that measures the goodness of fit of the model to the data. We assume uniform priors for all parameters except for a logarithmic prior for the Doppler $b$ parameter \citep{Youngblood2016}, and a Gaussian likelihood function. We used 100 walkers, ran for 50 autocorrelation times, and removed a burn-in period. Table~\ref{Tab:properties} lists selected best-fit parameters and Figure~\ref{fig:LyA_reconstruciton_results} shows the best fit and reconstructed line profiles and residuals.

\subsection{Fits to stellar \MgII\ k profiles and solar \Lya\ and \MgII\ k spectra} \label{subsec:Sun}

To enable direct comparison with our intrinsic \Lya\ profiles after correction for instrumental line broadening effects, we fit a modified Equation~\ref{eq:model} to our E230H \MgII\ k profiles (Figure~\ref{fig:MgII_data}, Table~\ref{Tab:properties_MgII}) and the average quiet Sun \Lya\ and \MgII\ k profiles shown in Figure~\ref{fig:solar_LyA_MgII} from \cite{Gunar2020,Gunar2021}. For these spectra, the ISM attenuation can be dropped from Equation~\ref{eq:model}. 

 The uneven peaks of the stellar \MgII\ and solar \Lya\ and \MgII\ profiles show that self-reversal asymmetry is present, which needs to be accommodated by an additional parameter in our symmetric model. We add an offset velocity parameter to $V_{\rm radial}$ in Equation~\ref{eq:em} called $V_{\rm reversal}$ that allows the self-reversal to be offset from the emission line, creating the asymmetry. This additional parameter is applied only in $\mathcal{V}^{\lambda}_{\rm norm}$, so that the velocity centroid of $\mathcal{V}^{\lambda}_{\rm norm}$ equals the sum of $V_{\rm radial}$ and $V_{\rm reversal}$. We did not include the $V_{\rm reversal}$ parameter for our \Lya\ reconstructions, because, except for possibly Kapteyn's Star, reliable information about the asymmetry of the \Lya\ line profiles cannot be inferred given the data quality and ISM attenuation. We fit the stellar \MgII\ and solar \Lya\ and \MgII\ profiles with the same fitting procedure as previously described. For the stellar \MgII\ lines we use the STIS LSFs. For the solar spectra, we assume Gaussian LSFs with FWHMs corresponding to the reported instrument resolutions: 0.08 \AA\ for IRIS (\MgII; \citealt{DePontieu2014}) and 0.086 \AA\ for SOHO/SUMER (\Lya; \citealt{Wilhelm1997}). Selected fit results are listed in Tables~\ref{Tab:properties} and \ref{Tab:properties_MgII} and shown in Figures~\ref{fig:MgII_data} and \ref{fig:LyA_MgII_comparison}.

The fitted solar \Lya\ and \MgII\ k $V_{\rm reversal}$ parameters are nearly identical (0.68 and 0.79 \kms, respectively), and we assume that the same holds true for the stellar spectra. To test the ability of our \Lya\ spectra to constrain $V_{\rm reversal}$ and the impact of fixing $V_{\rm reversal}$=0 \kms\ on our recovered parameters, we performed a \Lya\ reconstruction for HD 191408 where $V_{\rm reversal}$ was free to vary. We imposed a Gaussian prior of mean 0.44 \kms\ and standard deviation 0.10 \kms\ on $V_{\rm reversal}$ (i.e., that star's \MgII\ k median $V_{\rm reversal}$ value and 68\% confidence interval). The \Lya\ fit yielded a solution where the intrinsic \Lya\ flux decreased by $\sim$2\% and the ISM column density increased by 0.01 dex; both changes are similar to the 1-$\sigma$ uncertainties reported in Table~\ref{Tab:properties}. The $V_{\rm reversal}$ parameter median and 68\% confidence interval mirrored the prior exactly. Together, this shows that $V_{\rm reversal}$ is not uniquely constrained by our data and assuming symmetrical \Lya\ self-reversal has a small impact on the reconstructed fluxes.

\begin{deluxetable}{lcccccccc}
\tablecolumns{9}
\tablewidth{0pt}
\tablecaption{Selected \Lya\ Fitted Properties \label{Tab:properties}}
\tablehead{\colhead{Target} &
                  \colhead{$V_{\rm radial}$} &
                  \colhead{$V_{\rm HI}$} &
                  \colhead{log $N(HI)$} &
                  \colhead{$b_{\rm HI}$} &
                  \colhead{$p$} &
                  \colhead{$V_{\rm reversal}$} &
                  \colhead{$F$(\Lya)} &
                  \colhead{Peak-to-trough} \\
                  \colhead{Name} &
                  \colhead{(km s$^{-1}$)} &
                  \colhead{(km s$^{-1}$)} &
                  \colhead{cm$^{-2}$} & 
                  \colhead{(km s$^{-1}$)} &
                  \colhead{} &
                  \colhead{(km s$^{-1}$)} &
                  \colhead{(erg cm$^{-2}$ s$^{-1}$)} &
                  \colhead{ratio}
                  }
\startdata
82 Eri & 89.37$^{+2.43}_{-2.27}$ & 13.36$\pm$1.12 & 18.33$\pm$0.03 & 11.20$^{+0.99}_{-1.81}$  & 2.43$\pm$0.10 & =0 & (2.23$^{+0.07}_{-0.06}$) & 1.72$\pm$0.10   \\
     &  &  &  & &  & &  $\times$10$^{-12}$ &  \\
\hline
HD 191408 & -135.57$^{+2.18}_{-1.99}$ & -25.05$\pm$0.42 & 18.28$\pm$0.02 & 14.06$^{+0.46}_{-0.39}$ & 2.09$\pm$0.06 & =0 & (2.04$\pm$0.04) & 1.42$^{+0.05}_{-0.04}$ \\
     &  &  &  & & & & $\times$10$^{-12}$ &  \\
     \hline
Kapteyn's & 245.28$^{+0.80}_{-0.93}$ & -6.59$^{+21.70}_{-16.51}$ & 17.98$^{+0.36}_{-0.32}$ & =11.5 & 1.52$\pm$0.10 & =0 & (2.88$^{+0.16}_{-0.08}$) & 1.11$\pm$0.04  \\ 
Star  &  &  &  & & & &  $\times$10$^{-13}$ &   \\
\hline
GJ 411 & -87.16$^{+3.08}_{-3.04}$ & 3.91$^{+0.77}_{-0.74}$ & 17.84$\pm$0.03 & 11.74$\pm$0.49 & 1.50$^{+0.12}_{-0.13}$ & =0 & (1.62$^{+0.07}_{-0.06}$)  & 1.10$^{+0.05}_{-0.04}$ \\ 
     &  &  &  &  & & &  $\times$10$^{-12}$ &  \\
\hline
Barnard's & -109.66$^{+1.85}_{-1.75}$ & -23.84$^{+0.57}_{-0.53}$ & 17.72$\pm$0.03 & 10.79$^{+0.32}_{-0.33}$ & 1.27$\pm$0.10 & =0 & (5.83$\pm$0.02) & 1.03$\pm$0.02  \\
Star     &  &  &  & &  &  &   $\times$10$^{-13}$ &   \\
\hline \hline
Sun & -- & -- & -- & -- & 2.38$\pm$0.14 & 0.68$^{+0.82}_{-0.81}$ & -- & 1.67$^{+0.14}_{-0.13}$ \\
\enddata
\tablecomments{$V_{\rm radial}$ is the stellar radial velocity, $V_{\rm HI}$ is the ISM HI radial velocity, $N(HI)$ is the ISM column density, $b_{\rm HI}$ is the Doppler broadening parameter for the ISM absorbers, $p$ is the unitless self-absorption parameter, $V_{\rm reversal}$ is the offset velocity of the self-reversal, and $F$(\Lya) is the reconstructed stellar \Lya\ flux. All uncertainties represent the 68\% confidence interval, and the reported values are the median (values with an equal sign indicate the parameter was fixed at that value during the fit). HD 191408's large $b_{\rm HI}$ value indicates there are probably multiple interstellar clouds along this sightline. For the Sun, we only list the measured $p$ and peak-to-trough ratio values because $V_{\rm radial}$ is zero by definition in the heliocentric frame, no ISM attenuation is present, and the integrated flux of the \cite{Gunar2020} profile is not representative of the Sun's disk-integrated \Lya\ flux.}
\end{deluxetable}

\begin{deluxetable}{lcccc}
\tablecolumns{5}
\tablewidth{0pt}
\tablecaption{Selected \MgII\ k Fitted Properties \label{Tab:properties_MgII}}
\tablehead{\colhead{Target} &
                  \colhead{$V_{\rm radial}$} &
                  \colhead{$p$} &
                  \colhead{$V_{\rm reversal}$} &
                  \colhead{Peak-to-trough} \\
                  \colhead{Name} &
                  \colhead{(km s$^{-1}$)} &
                  \colhead{} &
                  \colhead{(km s$^{-1}$)} &
                  \colhead{ratio}
                  }
\startdata
82 Eri & 87.37$\pm$0.07 & 2.71$\pm$0.03 & 0.72$\pm$0.05 & 2.05$^{+0.04}_{-0.03}$  \\
     &  & & & \\
\hline
HD 191408 & -130.08$^{+0.14}_{-0.13}$ & 2.33$\pm$0.05 & 0.44$\pm$0.10 & 1.63$\pm$0.05 \\
     &  & & &  \\
     \hline
Kapteyn's & 242.46$^{+1.10}_{-1.17}$ & 0\tablenotemark{a} & $\sim$0\tablenotemark{a} & 1.00$^{+0.14}_{-0.00}$\\ 
Star  &  & &  &  \\
\hline
GJ 411 & -87.8$\pm$0.2 & 2.81$\pm$0.25 & 0.17$^{+0.14}_{-0.13}$ & 2.18$^{+0.39}_{-0.32}$  \\ 
     &  & &  & \\
\hline
Barnard's & -111.3$\pm$0.2 & 2.27$\pm$0.30 & -0.03$\pm$0.15 & 1.58$^{+0.30}_{-0.23}$ \\
Star     &   & &  &   \\
\hline \hline
Sun & -- & 2.59$\pm$0.04 & 0.79$\pm$0.07 & 1.91$\pm$0.05 \\
\enddata
\tablecomments{$V_{\rm radial}$ is the stellar radial velocity, $p$ is the unitless self reversal parameter, and $V_{\rm reversal}$ is the offset velocity centroid of the reversal from $V_{\rm radial}$. All uncertainties represent the 68\% confidence interval, and the reported values are the median.}
\tablenotetext{a}{Kapteyn's Star's self-reversal is poorly constrained due to the low signal-to-noise of the \MgII\ spectrum. We applied uniform priors to $p$ (0-3) and $V_{\rm reversal}$ (-0.3 to +0.3 km s$^{-1}$). The most likely value for $p$ is zero with a 1-$\sigma$ upper limit of 1.6, and $V_{\rm reversal}$ is unconstrained within the prior bounds.}
\end{deluxetable}

\subsection{Note on apparent \Lya\ flux evolution of Barnard's Star and Kapteyn's Star} \label{subsec:slit_losses}

The intrinsic \Lya\ fluxes we measure for Kapteyn's Star and Barnard's Star are significantly smaller than values reported in the literature measured from lower resolution spectra. Kapteyn's Star was observed with the COS G130M 2.5\arcsec\ PSA aperture on 2013-Sep-22, and \cite{Guinan2016} reported a \Lya\ flux 1.85$\times$ larger than that measured in this work using the STIS E140M 0.2\arcsec$\times$0.06\arcsec\ slit on 2019-Apr-03. Barnard's Star was observed on 2019-Mar-04 with the STIS G140M 52$\times$0.1\arcsec\ slit, and \cite{France2020} reported a \Lya\ flux 1.75$\times$ larger than that measured in this work with the STIS E140M 0.2\arcsec$\times$0.06\arcsec\ slit on 2018-Aug-12. In both cases, such a large difference cannot be attributed to differences in the ISM correction and may be astrophysical or instrumental in nature. Despite their old ages, Barnard's Star and Kapteyn's Star both exhibit chromospheric variability due to magnetic activity \citep{Guinan2016,France2020}, so activity cycles akin to the Sun's 11-year cycle could feasibly be responsible for these flux differences. However, STIS has known focus issues that noticeably affect the flux calibration of observations with narrow slit widths ($<$0.2\arcsec; \citealt{Proffitt2017,Riley2018}). All of our targets except GJ 411 were observed with the 0.06\arcsec\ slit for \Lya, meaning that the \Lya\ fluxes of Barnard's Star and Kapteyn's Star could be systematically low.

To shed further light on the cause of this flux discrepancy, we compare this work's \MgII\ flux for Barnard's Star (obtained with the STIS E230H 0.2\arcsec$\times$0.2\arcsec\ slit on 2018-Aug-12) to the \MgII\ flux that \cite{France2020} measured with the lower resolution STIS G230L 52\arcsec$\times$0.2\arcsec\ slit on 2019-Mar-04. Note that each \MgII\ measurement was taken on the same day as the \Lya\ measurement for that star. \cite{France2020} reported \MgII\ k \& h fluxes 3\% and 5\%, respectively, larger than our measured fluxes. Given the good agreement between the STIS G230L 52\arcsec$\times$0.2\arcsec\ and STIS E230H 0.2\arcsec$\times$0.2\arcsec\ \MgII\ fluxes of Barnard's Star taken seven months apart, we conclude that the large \Lya\ flux differences for Barnard's Star are attributable to STIS focus issues. 

We compare Kapteyn's Star's \MgII\ k \& h fluxes measured with the STIS E230H 0.2\arcsec$\times$0.2\arcsec\ slit on 2019-Apr-03 with the fluxes measured from IUE on 1987-Aug-19 presented in \cite{Guinan2016}. Kapteyn's Star's COS \Lya\ spectrum from 2013 does not have a contemporaneous \MgII\ spectrum. We find that Kapteyn's Star's \MgII\ flux is 15\% fainter in 2019 than in 1987, but given the low signal-to-noise of our STIS spectrum, the fluxes are consistent within the 95\% confidence interval. Therefore, it is plausible that the 85\% \Lya\ flux difference between 2013 and 2019 for Kapteyn's Star could be astrophysical or instrumental in nature. We caution the reader that the intrinsic \Lya\ fluxes reported for all stars except GJ 411, which was observed with the wider 0.2\arcsec$\times$0.2\arcsec\ slit, could be systematically low due to STIS flux calibration issues with narrow slits \citep{Proffitt2017,Riley2018}.

\section{Results} \label{sec:Results}

We find that self-reversal is present in the \Lya\ profiles of all five targets, and that the self-reversal depth is greater for earlier spectral types than for later spectral types. This is reflected in the finding that the best fit self-absorption parameter $p>$1 for all targets (Table~\ref{Tab:properties}). Barnard's Star has the smallest value $p$=1.27$\pm$0.10, and the other two M dwarfs (GJ 411 and Kapteyn's Star) have somewhat larger values ($p$=1.50$^{+0.12}_{-0.13}$ and $p$=1.52$\pm$0.10, respectively). The K dwarf HD 191408 has the next largest value ($p$=2.09$\pm$0.06), and the G dwarf 82 Eri has the largest at $p$=2.43$\pm$0.10. For comparison, we find that the average quiet Sun's \Lya\ profile has $p$=2.38$\pm$0.14, consistent with 82 Eri's.

Self-reversals in the \MgII\ profiles are apparent for all stars except Kapteyn's Star, which has a low signal-to-noise \MgII\ spectrum. However, unlike for \Lya, no trend between \MgII\ self-reversal and spectral type is seen among these five stars. The most likely value of Kapteyn's Star's \MgII\ $p$ is zero, but the 1-$\sigma$ upper limit is $p$=1.6, indicating that self-reversal cannot be confidently ruled out. The M dwarf Barnard's Star and the K dwarf HD 191408 have the next largest $p$ values and are equivalent within uncertainties ($p$=2.27$\pm$0.30 and $p$=2.33$\pm$0.05, respectively). The M dwarf GJ 411 and the G dwarf 82 Eri have the largest $p$ values and are roughly equivalent within uncertainties ($p$=2.81$\pm$0.25 and $p$=2.71$\pm$0.03, respectively). For comparison, we find that the average quiet Sun's \MgII\ k profile has $p$=2.59$\pm$0.04, significantly smaller than 82 Eri's. Except Kapteyn's Star, all stars including the Sun exhibit greater self-reversal in their \MgII\ profiles than in \Lya.

  \begin{figure}
         \centering
          \includegraphics[width=0.75\textwidth]{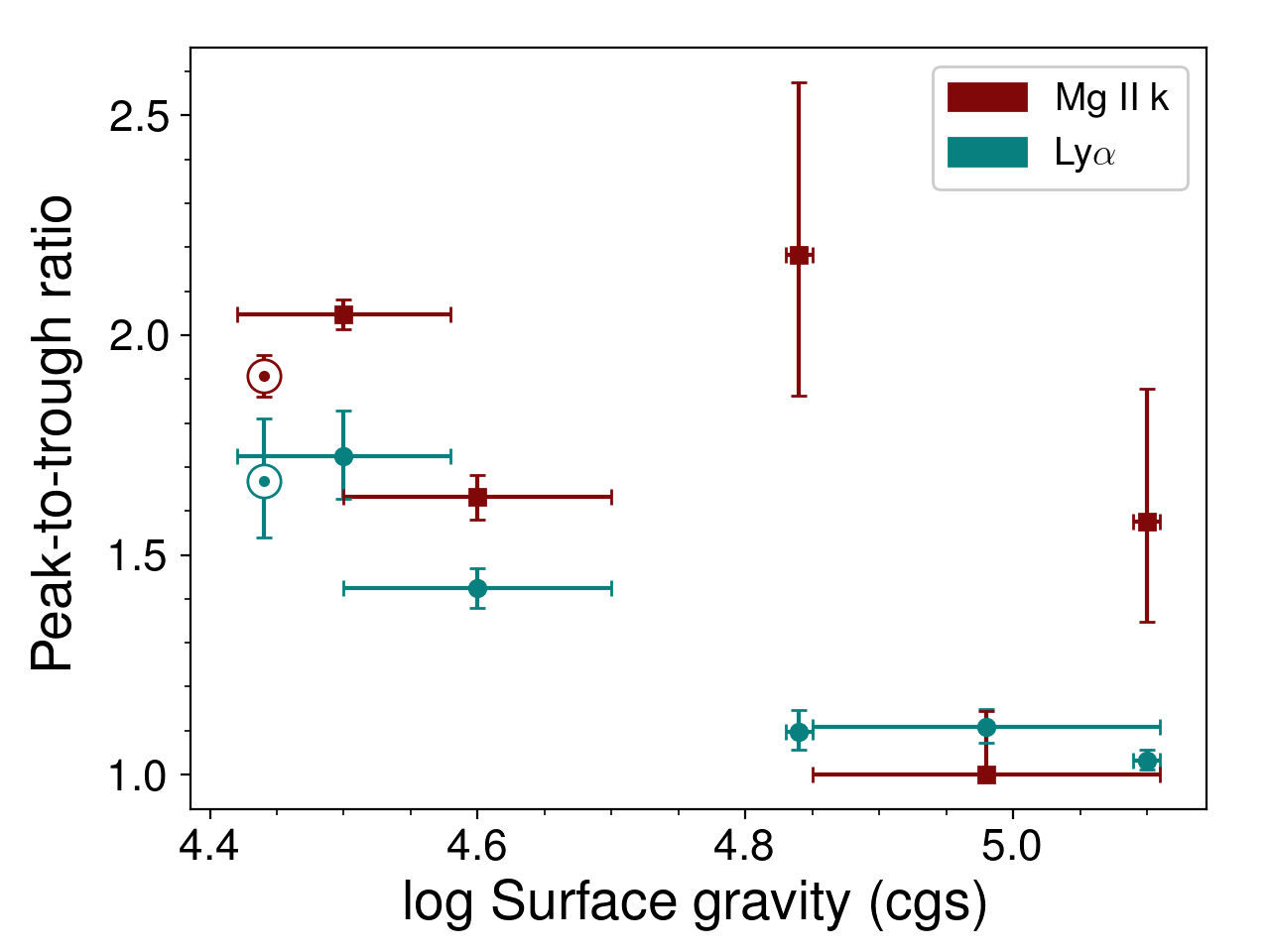}
          \caption{Peak-to-trough ratios for \Lya\ (teal points) and \MgII\ k (red points) are shown with 1-$\sigma$ uncertainties as a function of stellar surface gravity for the five targets. For the Sun (circle-dot symbols), we adopted log $g$ = 4.44.}
          \label{fig:logg_SR}
  \end{figure}

To further quantify self-reversal depth, we measure the peak-to-trough ratios of the intrinsic \Lya\ and \MgII\ profiles of our sample and the Sun. Unlike the self-absorption parameter $p$, this quantity can be measured directly from spectra and does not require fitting Equation~\ref{eq:em} to the data. However, we note that instrumental line broadening can decrease the measured peak-to-trough ratio and should be taken into account as appropriate. For example, instead of measuring the peak-to-trough ratio directly from the solar spectra shown in Figure~\ref{fig:solar_LyA_MgII}, we fit a model convolved with the instrumental LSF (Section~\ref{subsec:Sun}), allowing us to recover a un-convolved line profile from which to measure the peak-to-trough ratio. The peak-to-trough ratio is unity for profiles with no self-reversal, and it therefore does not distinguish between flat-topped lines (0$<p<$1) and pure Voigt profiles ($p$=0). For asymmetric profiles, the average flux density of the two peaks is used to compute the ratio. The median peak-to-trough ratios and 68\% confidence intervals determined from the ensemble MCMC samples (prior to convolution with the instrument LSFs) are reported in Tables~\ref{Tab:properties} and \ref{Tab:properties_MgII}.

We compare the \Lya\ and \MgII\ peak-to-trough ratios with surface gravity (Figure~\ref{fig:logg_SR}). We observe a clear decrease in \Lya\ peak-to-trough ratio from 1.7 to 1.0 with increasing surface gravity (log $g$ from 4.5 to 5.1), and the Sun follows this trend within uncertainties. For \MgII, the trend is not followed. The M dwarfs GJ 411 and Barnard's Star show self-reversals on par with the earlier type stars. Kapteyn's Star's \MgII\ spectrum has very low signal-to-noise, but our fits demonstrate that the peak-to-trough ratio is consistent with unity and a 1-$\sigma$ upper limit of 1.14. This is possible because the definition of the peak-to-trough ratio implies that any profile with $p<$1 will have a peak-to-trough ratio of unity; while Kapteyn's Star's \MgII\ $p$ value is highly uncertain (Table~\ref{Tab:properties_MgII}), 63\% of the self-absorption parameter $p$ parameter space is $<$1. We note that the results reflected in Figure~\ref{fig:logg_SR} remain similar when using the self-absorption parameter $p$ instead of peak-to-trough ratio.

\section{Discussion} \label{sec:Discussion}

Our results indicate that \Lya\ self-reversal depth correlates well with stellar surface gravity. This suggests that basic stellar structure plays a role in dictating the properties of the \Lya\ emitting regions of the upper chromosphere, perhaps most crucially the density. The results from lower-resolution spectroscopy of three high-radial velocity K and M dwarfs from \cite{Bourrier2017_Kepler444} and \cite{Schneider2019} follow this trend.

\MgII\ forms deeper in the atmosphere than \Lya\ (e.g., \citealt{Vernazza1981}), where densities are higher and perhaps less dependent on surface gravity effects. This could explain the absence of a trend with surface gravity for \MgII\ and the fact that \MgII\ consistently shows deeper self-reversals than \Lya. Notably, the \MgII\ line core shapes of the G-K dwarfs closely follow \Lya, but two of the M dwarf \MgII\ line cores exhibit very narrow and deep self-reversals in contrast to their \Lya\ profiles. This indicates major differences between the chromospheres of M dwarfs and earlier types, but could be influenced by other factors including the low activity and metallicity of our sample, and the lower signal-to-noise of the M dwarf \MgII\ spectra. The lower signal-to-noise translates into the large errors in Figure~\ref{fig:logg_SR}, which in turn may mask any correlation.

Based on spatially resolved \Lya\ and \MgII\ spectroscopy of the Sun, magnetic activity plays a role in controlling the self-reversal depth and asymmetry. For example, sunspots, plages, and network regions generally exhibit shallower self-reversals than quiet Sun regions (e.g., \citealt{Fontenla1988,Tian2009_LyALyB,Tian2009_transitionregion_sunspots,Schmit2015}). Measurements of activity and surface magnetism across the main sequence show clear differences between G, K, and M dwarfs. The average magnetic field flux (due to cancellation) across the quiet Sun is $\sim$1 Gauss, while sunspot regions have kilo-Gauss field fluxes. Global fields increase dramatically with decreasing stellar mass, reaching kilo-Gauss fluxes for M dwarfs (see \citealt{Donati2009} for a review). The greater surface magnetism of M dwarfs is likely a contributing factor to the self-reversal trend with spectral type that we observe, and it would be interesting to disentangle the effects of surface gravity and magnetism (or its proxy stellar activity) on the self-reversal of chromospheric emission lines.  A larger sample with a range of activity could allow for disentangling the specific effects of activity on self-reversal.

\subsection{The use of \MgII\ as a proxy for \Lya\ line core shape}

Our analysis of the average quiet Sun \Lya\ and \MgII\ profiles show that their asymmetries are nearly identical, but the solar \MgII\ self-reversal is deeper than \Lya's. Our stellar \Lya\ spectra do not provide strong constraints on the line asymmetries, but comparison with the unobstructed \MgII\ profiles show that the \MgII\ self-reversal is consistently deeper than for \Lya. The exception is Kapteyn's Star, but as previously noted, the quality of that star's \MgII\ data is poor. 

Our model tests in Section~\ref{sec:linefitting} show that the \MgII\ data cannot be successfully used directly as a template for the intrinsic \Lya\ profile during a reconstruction. However, \MgII\ spectra may still provide useful constraints on a \Lya\ reconstruction by providing an upper limit on the peak-to-trough ratio and an estimate for the asymmetry.

\subsection{The effect of ignoring self-reversal on \Lya\ reconstructions of low-radial velocity stars} \label{subsec:ignore}

The vast majority of nearby stars are more strongly affected by ISM attenuation than our target stars, and generally the line cores will not be observable. We use the intrinsic profiles derived in this work to test the accuracy of standard reconstruction techniques, which do not include self-reversals, for low radial velocity stars. We shifted the un-convolved intrinsic \Lya\ profiles from Figure~\ref{fig:LyA_MgII_comparison} to zero velocity, added ISM attenuation (see Equation~\ref{eq:abs}) with $V_{\rm HI}$= 0 \kms, $b_{\rm HI}$=11.5 \kms, and $N$(HI) = 10$^{18}$ cm$^{-2}$ (chosen to be typical of nearby sightlines, e.g., \citealt{Wood2005}), and convolved the profiles with the STIS E140M LSF for the 0.2\arcsec$\times$0.06\arcsec\ aperture. This grating and aperture setup was used for all of our stars except GJ 411, which used the 0.2\arcsec$\times$0.2\arcsec\ aperture, and achieves a spectral resolving power $R\approx$45,800. We gave the synthetic spectra flux uncertainties based on the original STIS data. Specifically, any flux density $F_{\lambda,i}$ in the synthetic spectrum would inherit the average errorbar value $E_{\lambda,i}$ of any flux density value near $F_{\lambda,i}$ in the original STIS E140M spectrum. Random noise, drawn from a normal distribution with mean zero and standard deviation equal to the error bar value, was added to the synthetic data. We also simulated STIS G140M spectra, because that mode is commonly used for G-M dwarf \Lya\ observations in recent HST cycles. For these simulated spectra, we used the STIS G140M LSF for the 52\arcsec$\times$0.2\arcsec\ aperture and adopted noise properties consistent with one orbit of exposure time for all five stars.

\begin{figure}
         \centering
          \includegraphics[width=0.65\textwidth]{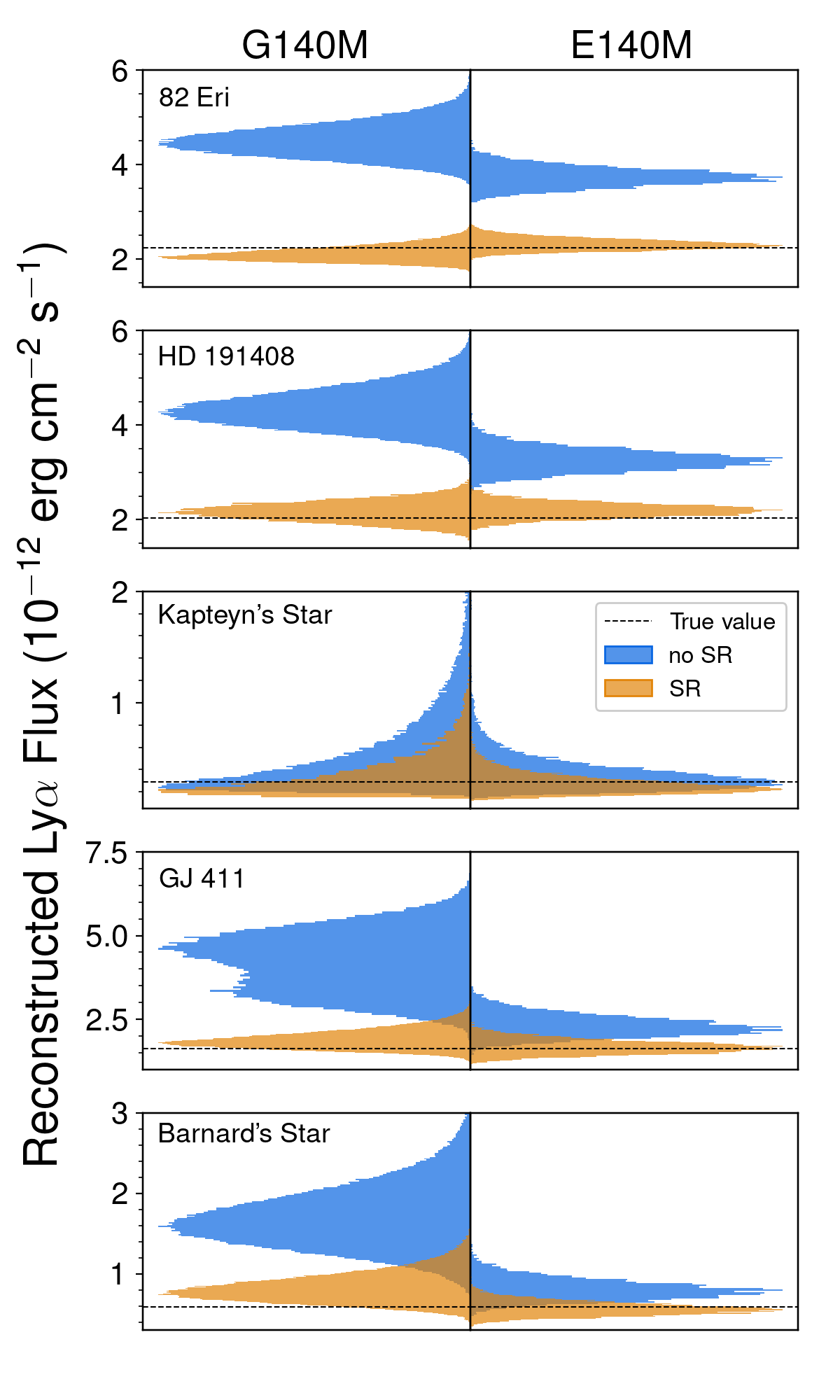}
          \caption{The histograms show the probability density for each intrinsic \Lya\ flux measured from the synthetic G140M and E140M spectra (left and right, respectively). The \Lya\ fluxes of the profiles used to make the synthetic data (shown by a dashed horizontal line) are compared to the fluxes recovered from the synthetic STIS data simulated for low-radial velocity stars (i.e., a case where the intrinsic stellar emission line and ISM absorption have the same velocity). The blue histograms shows the results where self-reversal (SR) was not allowed ($p$=0) and the orange histograms show fit results where self-reversal was restricted to a realistic range of $p$ based on the stellar surface gravity (see text for specific ranges).}
          \label{fig:fake_g140m_e140m}
  \end{figure}

We fit each synthetic spectrum twice with different restrictions on the self-absorption parameter $p$: $p$=0 (i.e., no reversal) and restricted $p$ to realistic ranges based on the stellar surface gravity. For 82 Eri, 1.8$\leq p \leq$2.8, for HD 191408, 1.5$\leq p \leq$2.5, and for the three M dwarfs, 1.0$\leq p \leq$1.8. The resulting intrinsic \Lya\ fluxes compared with the original values presented in Table~\ref{Tab:properties} are shown in Figure~\ref{fig:fake_g140m_e140m}.

We find that including self-reversal is generally essential to recover the true \Lya\ flux for either grating. For 82 Eri and HD 191408, the reconstructions without self-reversal overestimate the true flux between 60\% and 110\%, and including self-reversal brings agreement with the input flux to within 7\%. The M dwarf G140M reconstructions with no self-reversal can overestimate \Lya\ fluxes by 52\%-185\%, levels comparable to or greater than the earlier type stars, but this overestimate is less severe for the E140M spectra (13\%-40\%). Including self-reversal for the M dwarfs significantly improves agreement to within the 2\%-15\% of the true flux. The exception is Barnard's Star, whose G140M reconstruction with self-reversal overestimates the input flux by 38\%. Kapteyn's Star's simulated spectra are very noisy, and while reconstructions with or without self-reversal agree with the true flux within the 68\% confidence interval, including self-reversal narrows the probability distribution to be tighter around the true flux.

For context in interpreting these results, the integrated flux of a Voigt profile (i.e., Equation~\ref{eq:em} with $p$=0) will be larger than the integrated flux of a flat-topped or self-reversed Voigt profile (i.e., Equation~\ref{eq:em} with $p>$0) by about 85\% for $p$=1, 150\% for $p$=1.5, 220\% for $p$=2, and 305\% for $p$=2.5. How much the intrinsic flux is overestimated when ignoring self-reversal depends mainly on if increasing the ISM column density can bring the attenuated Voigt profile into good agreement with the data. With lower spectral resolution and/or lower signal-to-noise, this becomes much easier for the fitting routine to accomplish. For broad intrinsic lines with deep self-reversal like the G and K dwarfs, simply increasing the ISM attenuation is not sufficient for agreement with the data. The fits must also compromise by increasing the Voigt widths and decreasing the amplitude, and this decreases the overestimation relative to the M dwarfs. However, the quality of the G and K dwarf reconstructions without self-reversal is clearly poorer, indicating that the model is misspecified (i.e., self-reversal is needed).

Self-reversal is generally neglected for M dwarf reconstructions in the literature \citep{Youngblood2016,Bourrier2018,dosSantos2020,Linsky2020,Youngblood2021}, and the current results indicate that revisiting these past reconstructions with more realistic constraints on self-reversal would be worthwhile. Additional \Lya\ observations of high-radial velocity stars could better inform the most likely parameter space for self-reversal depth and asymmetry, especially its dependence on stellar activity and metallicity. Our sample of five stars is small and, given that stars with large space velocities tend to be old, is skewed to lower activity and metallicity. Spatially-resolved solar \Lya\ observations show that greater surface activity lessens self-reversal (e.g., \citealt{Fontenla1988}), so more active stars likely exhibit shallower self-reversals than the stars in this sample. Thus, the \Lya\ flux overestimates presented in this section may be a worst-case scenario. 

\subsection{Recommendations for exoplanet photochemical studies that rely on \Lya\ fluxes} \label{subsec:recommendations}

A host star's UV spectrum is a required input for exoplanet photochemical models, which are used to understand the chemical makeup of exoplanet atmospheres and compare to or predict observations (e.g., \citealt{MillerRicciKempton2012,Rugheimer2015,Miguel2015,Arney2017}). The \Lya\ emission line is an order of magnitude brighter than other lines in the far-UV spectra of late type stars, especially M dwarfs \citep{France2013}, and is thus important to accurately capture. Although much effort has been expended in obtaining \Lya\ fluxes unaffected by the ISM (e.g., \citealt{Wood2005,Youngblood2016,Schneider2019}), uncertainties in the reconstructed fluxes remain and are rarely accounted for in photochemical models (Teal et al. under review). Authors who use \Lya\ fluxes should include caveats with their assessments and run models with a range of fluxes consistent with potential systematic effects and the uncertainty in the interstellar absorption toward their particular target. In addition to the systematic effects described in Section~\ref{subsec:ignore} related to reconstructions that neglect self-reversal, this paper also notes large flux changes over the years that may be due to either STIS flux calibration issues with narrow slits or stellar activity cycles (Section~\ref{subsec:slit_losses}). Finally, any particular star's reported \Lya\ fluxes may not be that far off systematically, because earlier spectral types are more affected by self-reversal than later types and highly active stars may be minimally affected. As new observations of planetary atmospheres exhibiting disequilibrium chemistry become available with JWST and other upcoming facilities, the urgency of characterizing systematic uncertainties affecting host star \Lya\ fluxes will increase.

\section{Conclusions} \label{sec:Conclusions}

New UV spectra from the \emph{Hubble Space Telescope} STIS instrument of five high radial velocity G, K, and M dwarfs indicate that most main sequence stars likely exhibit self-reversal in their \Lya\ line profiles. This result has implications for models of stellar upper atmospheres, interstellar medium measurements, and exoplanet atmosphere photochemical models. We show that basic stellar structure, as traced by surface gravity, appears to be the most important indicator for \Lya\ self-reversal; stars with greater surface gravity (lower mass) display weaker self-reversals. However, \MgII\ self-reversals, which always appear to be deeper than  \Lya\ self-reversals, do not appear to follow this trend. Nonetheless, commonalities exist between the \Lya\ and \MgII\ self-reversals, and we discuss how high-resolution \MgII\ spectra can provide {\em a priori} estimates of \Lya\ self-reversals. We show that neglecting self-reversal from \Lya\ reconstructions using STIS E140M spectra can lead to flux overestimations of up to 70\% for G and K dwarfs and as much as 40\% for M dwarfs. With the lower resolution G140M grating, neglecting self-reversal worsens the overestimates to up to 110\% for G and K dwarfs and up to 180\% for M dwarfs. Stars more magnetically active than those in our sample are likely less affected by self-reversal. These results can improve the accuracy of \Lya\ reconstructions for the vast majority of stars whose \Lya\ cores are obstructed by interstellar gas and inform models of stellar chromospheres.

\acknowledgments
We thank the anonymous referee for comments that improved the quality of the paper. A.Y. acknowledges support by an appointment to the NASA Postdoctoral Program at Goddard Space Flight Center, administered by USRA through a contract with NASA. A.Y. and J.L. thank the Space Telescope Science Institute for support through grant HST-GO-15190. This research is based on observations made with the NASA/ESA \emph{Hubble Space Telescope} obtained from the Mikulski Archive for Space Telescopes (MAST) at the Space Telescope Science Institute, which is operated by the Association of Universities for Research in Astronomy, Inc., under NASA contract NAS 5–26555. These observations are associated with program HST-GO-15190 (PI: A. Youngblood). The specific observations analyzed can be accessed via \dataset[https://doi.org/10.17909/t9-6dey-ww08]{https://doi.org/10.17909/t9-6dey-ww08}. A.Y. thanks Aki Roberge for commenting on the paper and Joleen Carlberg and Crystal Mannfolk for their role in executing the observing program at STScI.
This research has made use of the NASA Exoplanet Archive, which is operated by the California Institute of Technology, under contract with the National Aeronautics and Space Administration under the Exoplanet Exploration Program. This research has made use of NASA's Astrophysics Data System Bibliographic Services and the SIMBAD database, operated at CDS, Strasbourg, France.

\facilities{HST}
\software{Astropy \citep{Robitaille2013}, IPython \citep{Perez2007}, Matplotlib \citep{Hunter2007}, NumPy and SciPy \citep{VanderWalt2011}, lyapy \citep{Youngblood2016}, emcee \citep{Foreman-Mackey2013}.}

\bibliography{main.bbl}{}

\begin{thebibliography}{}
\expandafter\ifx\csname natexlab\endcsname\relax\def\natexlab#1{#1}\fi
\providecommand{\url}[1]{\href{#1}{#1}}
\providecommand{\dodoi}[1]{doi:~\href{http://doi.org/#1}{\nolinkurl{#1}}}
\providecommand{\doeprint}[1]{\href{http://ascl.net/#1}{\nolinkurl{http://ascl.net/#1}}}
\providecommand{\doarXiv}[1]{\href{https://arxiv.org/abs/#1}{\nolinkurl{https://arxiv.org/abs/#1}}}

\bibitem[{{Anglada-Escude} {et~al.}(2014){Anglada-Escude}, {Arriagada},
  {Tuomi}, {Zechmeister}, {Jenkins}, {Ofir}, {Dreizler}, {Gerlach}, {Marvin},
  {Reiners}, {Jeffers}, {Butler}, {Vogt}, {Amado}, {Rodriguez-Lopez},
  {Berdinas}, {Morin}, {Crane}, {Shectman}, {Thompson}, {Diaz}, {Rivera},
  {Sarmiento}, \& {Jones}}]{Anglada-Escude2014}
{Anglada-Escude}, G., {Arriagada}, P., {Tuomi}, M., {et~al.} 2014, \mnras, 443,
  L89, \dodoi{10.1093/mnrasl/slu076}

\bibitem[{{Arney} {et~al.}(2017){Arney}, {Meadows}, {Domagal-Goldman},
  {Deming}, {Robinson}, {Tovar}, {Wolf}, \& {Schwieterman}}]{Arney2017}
{Arney}, G.~N., {Meadows}, V.~S., {Domagal-Goldman}, S.~D., {et~al.} 2017,
  \apj, 836, 49, \dodoi{10.3847/1538-4357/836/1/49}

\bibitem[{{Ayres}(1979)}]{Ayres1979}
{Ayres}, T.~R. 1979, \apj, 228, 509, \dodoi{10.1086/156873}

\bibitem[{{Bidelman}(1985)}]{Bidelman1985}
{Bidelman}, W.~P. 1985, \apjs, 59, 197, \dodoi{10.1086/191069}

\bibitem[{{Bourrier} {et~al.}(2017){Bourrier}, {Ehrenreich}, {Allart},
  {Wyttenbach}, {Semaan}, {Astudillo-Defru}, {Gracia-Bern{\'a}}, {Lovis},
  {Pepe}, {Thomas}, \& {Udry}}]{Bourrier2017_Kepler444}
{Bourrier}, V., {Ehrenreich}, D., {Allart}, R., {et~al.} 2017, \aap, 602, A106,
  \dodoi{10.1051/0004-6361/201730542}

\bibitem[{{Bourrier} {et~al.}(2018){Bourrier}, {Lecavelier des Etangs},
  {Ehrenreich}, {Sanz-Forcada}, {Allart}, {Ballester}, {Buchhave}, {Cohen},
  {Deming}, {Evans}, {Garc{\'\i}a Mu{\~n}oz}, {Henry}, {Kataria}, {Lavvas},
  {Lewis}, {L{\'o}pez-Morales}, {Marley}, {Sing}, \& {Wakeford}}]{Bourrier2018}
{Bourrier}, V., {Lecavelier des Etangs}, A., {Ehrenreich}, D., {et~al.} 2018,
  \aap, 620, A147, \dodoi{10.1051/0004-6361/201833675}

\bibitem[{{Boyajian} {et~al.}(2012){Boyajian}, {von Braun}, {van Belle},
  {McAlister}, {ten Brummelaar}, {Kane}, {Muirhead}, {Jones}, {White},
  {Schaefer}, {Ciardi}, {Henry}, {L{\'o}pez-Morales}, {Ridgway}, {Gies}, {Jao},
  {Rojas-Ayala}, {Parks}, {Sturmann}, {Sturmann}, {Turner}, {Farrington},
  {Goldfinger}, \& {Berger}}]{Boyajian2012}
{Boyajian}, T.~S., {von Braun}, K., {van Belle}, G., {et~al.} 2012, \apj, 757,
  112, \dodoi{10.1088/0004-637X/757/2/112}

\bibitem[{{Carleo} {et~al.}(2021){Carleo}, {Youngblood}, {Redfield}, {Casasayas
  Barris}, {Ayres}, {Vannier}, {Fossati}, {Palle}, {Livingston}, {Lanza},
  {Niraula}, {Alvarado-G{\'o}mez}, {Chen}, {Gandolfi}, {Guenther}, {Linsky},
  {Nagel}, {Narita}, {Nortmann}, {Shkolnik}, \& {Stangret}}]{Carleo2021}
{Carleo}, I., {Youngblood}, A., {Redfield}, S., {et~al.} 2021, \aj, 161, 136,
  \dodoi{10.3847/1538-3881/abdb2f}

\bibitem[{Cowan \& Dieke(1948)}]{Cowan1948}
Cowan, R.~D., \& Dieke, G.~H. 1948, Rev. Mod. Phys., 20, 418,
  \dodoi{10.1103/RevModPhys.20.418}

\bibitem[{{Cutri} {et~al.}(2003){Cutri}, {Skrutskie}, {van Dyk}, {Beichman},
  {Carpenter}, {Chester}, {Cambresy}, {Evans}, {Fowler}, {Gizis}, {Howard},
  {Huchra}, {Jarrett}, {Kopan}, {Kirkpatrick}, {Light}, {Marsh}, {McCallon},
  {Schneider}, {Stiening}, {Sykes}, {Weinberg}, {Wheaton}, {Wheelock}, \&
  {Zacarias}}]{Cutri2003}
{Cutri}, R.~M., {Skrutskie}, M.~F., {van Dyk}, S., {et~al.} 2003, {2MASS All
  Sky Catalog of point sources.}

\bibitem[{{De Pontieu} {et~al.}(2014){De Pontieu}, {Title}, {Lemen}, {Kushner},
  {Akin}, {Allard}, {Berger}, {Boerner}, {Cheung}, {Chou}, {Drake}, {Duncan},
  {Freeland}, {Heyman}, {Hoffman}, {Hurlburt}, {Lindgren}, {Mathur}, {Rehse},
  {Sabolish}, {Seguin}, {Schrijver}, {Tarbell}, {W{\"u}lser}, {Wolfson},
  {Yanari}, {Mudge}, {Nguyen-Phuc}, {Timmons}, {van Bezooijen}, {Weingrod},
  {Brookner}, {Butcher}, {Dougherty}, {Eder}, {Knagenhjelm}, {Larsen},
  {Mansir}, {Phan}, {Boyle}, {Cheimets}, {DeLuca}, {Golub}, {Gates}, {Hertz},
  {McKillop}, {Park}, {Perry}, {Podgorski}, {Reeves}, {Saar}, {Testa}, {Tian},
  {Weber}, {Dunn}, {Eccles}, {Jaeggli}, {Kankelborg}, {Mashburn}, {Pust},
  {Springer}, {Carvalho}, {Kleint}, {Marmie}, {Mazmanian}, {Pereira}, {Sawyer},
  {Strong}, {Worden}, {Carlsson}, {Hansteen}, {Leenaarts}, {Wiesmann},
  {Aloise}, {Chu}, {Bush}, {Scherrer}, {Brekke}, {Martinez-Sykora}, {Lites},
  {McIntosh}, {Uitenbroek}, {Okamoto}, {Gummin}, {Auker}, {Jerram}, {Pool}, \&
  {Waltham}}]{DePontieu2014}
{De Pontieu}, B., {Title}, A.~M., {Lemen}, J.~R., {et~al.} 2014, \solphys, 289,
  2733, \dodoi{10.1007/s11207-014-0485-y}

\bibitem[{{D{\'\i}az} {et~al.}(2019){D{\'\i}az}, {Delfosse}, {Hobson},
  {Boisse}, {Astudillo-Defru}, {Bonfils}, {Henry}, {Arnold}, {Bouchy},
  {Bourrier}, {Brugger}, {Dalal}, {Deleuil}, {Demangeon}, {Dolon}, {Dumusque},
  {Forveille}, {Hara}, {H{\'e}brard}, {Kiefer}, {Lopez}, {Mignon}, {Moreau},
  {Mousis}, {Moutou}, {Pepe}, {Perruchot}, {Richaud}, {Santerne}, {Santos},
  {Sottile}, {Stalport}, {S{\'e}gransan}, {Udry}, {Unger}, \&
  {Wilson}}]{Diaz2019}
{D{\'\i}az}, R.~F., {Delfosse}, X., {Hobson}, M.~J., {et~al.} 2019, \aap, 625,
  A17, \dodoi{10.1051/0004-6361/201935019}

\bibitem[{{Donati} \& {Landstreet}(2009)}]{Donati2009}
{Donati}, J.~F., \& {Landstreet}, J.~D. 2009, \araa, 47, 333,
  \dodoi{10.1146/annurev-astro-082708-101833}

\bibitem[{{dos Santos} {et~al.}(2020){dos Santos}, {Ehrenreich}, {Bourrier},
  {Astudillo-Defru}, {Bonfils}, {Forget}, {Lovis}, {Pepe}, \&
  {Udry}}]{dosSantos2020}
{dos Santos}, L.~A., {Ehrenreich}, D., {Bourrier}, V., {et~al.} 2020, \aap,
  634, L4, \dodoi{10.1051/0004-6361/201937327}

\bibitem[{{Ducati}(2002)}]{Ducati2002}
{Ducati}, J.~R. 2002, VizieR Online Data Catalog

\bibitem[{{Eker} {et~al.}(2018){Eker}, {Bak{\i}{\c{s}}}, {Bilir}, {Soydugan},
  {Steer}, {Soydugan}, {Bak{\i}{\c{s}}}, {Ali{\c{c}}avu{\c{s}}}, {Aslan}, \&
  {Alpsoy}}]{Eker2018}
{Eker}, Z., {Bak{\i}{\c{s}}}, V., {Bilir}, S., {et~al.} 2018, \mnras, 479,
  5491, \dodoi{10.1093/mnras/sty1834}

\bibitem[{{Emerich} {et~al.}(2005){Emerich}, {Lemaire}, {Vial}, {Curdt},
  {Sch{\"u}hle}, \& {Wilhelm}}]{Emerich2005}
{Emerich}, C., {Lemaire}, P., {Vial}, J.-C., {et~al.} 2005, \icarus, 178, 429,
  \dodoi{10.1016/j.icarus.2005.05.002}

\bibitem[{{Feng} {et~al.}(2017){Feng}, {Tuomi}, \& {Jones}}]{Feng2017}
{Feng}, F., {Tuomi}, M., \& {Jones}, H.~R.~A. 2017, \aap, 605, A103,
  \dodoi{10.1051/0004-6361/201730406}

\bibitem[{{Fontenla} {et~al.}(1988){Fontenla}, {Reichmann}, \&
  {Tandberg-Hanssen}}]{Fontenla1988}
{Fontenla}, J., {Reichmann}, E.~J., \& {Tandberg-Hanssen}, E. 1988, \apj, 329,
  464, \dodoi{10.1086/166392}

\bibitem[{{Fontenla} {et~al.}(2016){Fontenla}, {Linsky}, {Witbrod}, {France},
  {Buccino}, {Mauas}, {Vieytes}, \& {Walkowicz}}]{Fontenla2016}
{Fontenla}, J.~M., {Linsky}, J.~L., {Witbrod}, J., {et~al.} 2016, \apj, 830,
  154, \dodoi{10.3847/0004-637X/830/2/154}

\bibitem[{{Forbes} \& {Bridges}(2010)}]{Forbes2010}
{Forbes}, D.~A., \& {Bridges}, T. 2010, \mnras, 404, 1203,
  \dodoi{10.1111/j.1365-2966.2010.16373.x}

\bibitem[{Foreman-Mackey {et~al.}(2013)Foreman-Mackey, Hogg, Lang, \&
  Goodman}]{Foreman-Mackey2013}
Foreman-Mackey, D., Hogg, D.~W., Lang, D., \& Goodman, J. 2013, Publications of
  the Astronomical Society of the Pacific, 125, 306, \dodoi{10.1086/670067}

\bibitem[{France {et~al.}(2013)France, Froning, Linsky, Roberge, Stocke, Tian,
  Bushinsky, D{\'{e}}sert, Mauas, Vieytes, \& Walkowicz}]{France2013}
France, K., Froning, C.~S., Linsky, J.~L., {et~al.} 2013, The Astrophysical
  Journal, 763, 149, \dodoi{10.1088/0004-637X/763/2/149}

\bibitem[{{France} {et~al.}(2020){France}, {Duvvuri}, {Egan}, {Koskinen},
  {Wilson}, {Youngblood}, {Froning}, {Brown}, {Alvarado-G{\'o}mez},
  {Berta-Thompson}, {Drake}, {Garraffo}, {Kaltenegger}, {Kowalski}, {Linsky},
  {Loyd}, {Mauas}, {Miguel}, {Pineda}, {Rugheimer}, {Schneider}, {Tian}, \&
  {Vieytes}}]{France2020}
{France}, K., {Duvvuri}, G., {Egan}, H., {et~al.} 2020, \aj, 160, 237,
  \dodoi{10.3847/1538-3881/abb465}

\bibitem[{{Gagn{\'e}} {et~al.}(2018){Gagn{\'e}}, {Mamajek}, {Malo}, {Riedel},
  {Rodriguez}, {Lafreni{\`e}re}, {Faherty}, {Roy-Loubier}, {Pueyo}, {Robin}, \&
  {Doyon}}]{Gagne2018}
{Gagn{\'e}}, J., {Mamajek}, E.~E., {Malo}, L., {et~al.} 2018, \apj, 856, 23,
  \dodoi{10.3847/1538-4357/aaae09}

\bibitem[{{Gaia Collaboration} {et~al.}(2016){Gaia Collaboration}, {Prusti},
  {de Bruijne}, {Brown}, {Vallenari}, {Babusiaux}, {Bailer-Jones}, {Bastian},
  {Biermann}, {Evans}, {Eyer}, {Jansen}, {Jordi}, {Klioner}, {Lammers},
  {Lindegren}, {Luri}, {Mignard}, {Milligan}, {Panem}, {Poinsignon},
  {Pourbaix}, {Randich}, {Sarri}, {Sartoretti}, {Siddiqui}, {Soubiran},
  {Valette}, {van Leeuwen}, {Walton}, {Aerts}, {Arenou}, {Cropper}, {Drimmel},
  {H{\o}g}, {Katz}, {Lattanzi}, {O'Mullane}, {Grebel}, {Holland}, {Huc},
  {Passot}, {Bramante}, {Cacciari}, {Casta{\~n}eda}, {Chaoul}, {Cheek}, {De
  Angeli}, {Fabricius}, {Guerra}, {Hern{\'a}ndez}, {Jean-Antoine-Piccolo},
  {Masana}, {Messineo}, {Mowlavi}, {Nienartowicz}, {Ord{\'o}{\~n}ez-Blanco},
  {Panuzzo}, {Portell}, {Richards}, {Riello}, {Seabroke}, {Tanga},
  {Th{\'e}venin}, {Torra}, {Els}, {Gracia-Abril}, {Comoretto},
  {Garcia-Reinaldos}, {Lock}, {Mercier}, {Altmann}, {Andrae}, {Astraatmadja},
  {Bellas-Velidis}, {Benson}, {Berthier}, {Blomme}, {Busso}, {Carry},
  {Cellino}, {Clementini}, {Cowell}, {Creevey}, {Cuypers}, {Davidson}, {De
  Ridder}, {de Torres}, {Delchambre}, {Dell'Oro}, {Ducourant}, {Fr{\'e}mat},
  {Garc{\'\i}a-Torres}, {Gosset}, {Halbwachs}, {Hambly}, {Harrison}, {Hauser},
  {Hestroffer}, {Hodgkin}, {Huckle}, {Hutton}, {Jasniewicz}, {Jordan},
  {Kontizas}, {Korn}, {Lanzafame}, {Manteiga}, {Moitinho}, {Muinonen},
  {Osinde}, {Pancino}, {Pauwels}, {Petit}, {Recio-Blanco}, {Robin}, {Sarro},
  {Siopis}, {Smith}, {Smith}, {Sozzetti}, {Thuillot}, {van Reeven}, {Viala},
  {Abbas}, {Abreu Aramburu}, {Accart}, {Aguado}, {Allan}, {Allasia},
  {Altavilla}, {{\'A}lvarez}, {Alves}, {Anderson}, {Andrei}, {Anglada Varela},
  {Antiche}, {Antoja}, {Ant{\'o}n}, {Arcay}, {Atzei}, {Ayache}, {Bach},
  {Baker}, {Balaguer-N{\'u}{\~n}ez}, {Barache}, {Barata}, {Barbier}, {Barblan},
  {Baroni}, {Barrado y Navascu{\'e}s}, {Barros}, {Barstow}, {Becciani},
  {Bellazzini}, {Bellei}, {Bello Garc{\'\i}a}, {Belokurov}, {Bendjoya},
  {Berihuete}, {Bianchi}, {Bienaym{\'e}}, {Billebaud}, {Blagorodnova},
  {Blanco-Cuaresma}, {Boch}, {Bombrun}, {Borrachero}, {Bouquillon}, {Bourda},
  {Bouy}, {Bragaglia}, {Breddels}, {Brouillet}, {Br{\"u}semeister},
  {Bucciarelli}, {Budnik}, {Burgess}, {Burgon}, {Burlacu}, {Busonero}, {Buzzi},
  {Caffau}, {Cambras}, {Campbell}, {Cancelliere}, {Cantat-Gaudin}, {Carlucci},
  {Carrasco}, {Castellani}, {Charlot}, {Charnas}, {Charvet}, {Chassat},
  {Chiavassa}, {Clotet}, {Cocozza}, {Collins}, {Collins}, {Costigan}, {Crifo},
  {Cross}, {Crosta}, {Crowley}, {Dafonte}, {Damerdji}, {Dapergolas}, {David},
  {David}, {De Cat}, {de Felice}, {de Laverny}, {De Luise}, {De March}, {de
  Martino}, {de Souza}, {Debosscher}, {del Pozo}, {Delbo}, {Delgado},
  {Delgado}, {di Marco}, {Di Matteo}, {Diakite}, {Distefano}, {Dolding}, {Dos
  Anjos}, {Drazinos}, {Dur{\'a}n}, {Dzigan}, {Ecale}, {Edvardsson}, {Enke},
  {Erdmann}, {Escolar}, {Espina}, {Evans}, {Eynard Bontemps}, {Fabre},
  {Fabrizio}, {Faigler}, {Falc{\~a}o}, {Farr{\`a}s Casas}, {Faye}, {Federici},
  {Fedorets}, {Fern{\'a}ndez-Hern{\'a}ndez}, {Fernique}, {Fienga}, {Figueras},
  {Filippi}, {Findeisen}, {Fonti}, {Fouesneau}, {Fraile}, {Fraser}, {Fuchs},
  {Furnell}, {Gai}, {Galleti}, {Galluccio}, {Garabato}, {Garc{\'\i}a-Sedano},
  {Gar{\'e}}, {Garofalo}, {Garralda}, {Gavras}, {Gerssen}, {Geyer}, {Gilmore},
  {Girona}, {Giuffrida}, {Gomes}, {Gonz{\'a}lez-Marcos},
  {Gonz{\'a}lez-N{\'u}{\~n}ez}, {Gonz{\'a}lez-Vidal}, {Granvik}, {Guerrier},
  {Guillout}, {Guiraud}, {G{\'u}rpide}, {Guti{\'e}rrez-S{\'a}nchez}, {Guy},
  {Haigron}, {Hatzidimitriou}, {Haywood}, {Heiter}, {Helmi}, {Hobbs},
  {Hofmann}, {Holl}, {Holland}, {Hunt}, {Hypki}, {Icardi}, {Irwin}, {Jevardat
  de Fombelle}, {Jofr{\'e}}, {Jonker}, {Jorissen}, {Julbe}, {Karampelas},
  {Kochoska}, {Kohley}, {Kolenberg}, {Kontizas}, {Koposov}, {Kordopatis},
  {Koubsky}, {Kowalczyk}, {Krone-Martins}, {Kudryashova}, {Kull}, {Bachchan},
  {Lacoste-Seris}, {Lanza}, {Lavigne}, {Le Poncin-Lafitte}, {Lebreton},
  {Lebzelter}, {Leccia}, {Leclerc}, {Lecoeur-Taibi}, {Lemaitre}, {Lenhardt},
  {Leroux}, {Liao}, {Licata}, {Lindstr{\o}m}, {Lister}, {Livanou}, {Lobel},
  {L{\"o}ffler}, {L{\'o}pez}, {Lopez-Lozano}, {Lorenz}, {Loureiro},
  {MacDonald}, {Magalh{\~a}es Fernandes}, {Managau}, {Mann}, {Mantelet},
  {Marchal}, {Marchant}, {Marconi}, {Marie}, {Marinoni}, {Marrese},
  {Marschalk{\'o}}, {Marshall}, {Mart{\'\i}n-Fleitas}, {Martino}, {Mary},
  {Matijevi{\v{c}}}, {Mazeh}, {McMillan}, {Messina}, {Mestre}, {Michalik},
  {Millar}, {Miranda}, {Molina}, {Molinaro}, {Molinaro}, {Moln{\'a}r},
  {Moniez}, {Montegriffo}, {Monteiro}, {Mor}, {Mora}, {Morbidelli}, {Morel},
  {Morgenthaler}, {Morley}, {Morris}, {Mulone}, {Muraveva}, {Musella},
  {Narbonne}, {Nelemans}, {Nicastro}, {Noval}, {Ord{\'e}novic},
  {Ordieres-Mer{\'e}}, {Osborne}, {Pagani}, {Pagano}, {Pailler}, {Palacin},
  {Palaversa}, {Parsons}, {Paulsen}, {Pecoraro}, {Pedrosa}, {Pentik{\"a}inen},
  {Pereira}, {Pichon}, {Piersimoni}, {Pineau}, {Plachy}, {Plum}, {Poujoulet},
  {Pr{\v{s}}a}, {Pulone}, {Ragaini}, {Rago}, {Rambaux}, {Ramos-Lerate},
  {Ranalli}, {Rauw}, {Read}, {Regibo}, {Renk}, {Reyl{\'e}}, {Ribeiro},
  {Rimoldini}, {Ripepi}, {Riva}, {Rixon}, {Roelens}, {Romero-G{\'o}mez},
  {Rowell}, {Royer}, {Rudolph}, {Ruiz-Dern}, {Sadowski}, {Sagrist{\`a}
  Sell{\'e}s}, {Sahlmann}, {Salgado}, {Salguero}, {Sarasso}, {Savietto},
  {Schnorhk}, {Schultheis}, {Sciacca}, {Segol}, {Segovia}, {Segransan},
  {Serpell}, {Shih}, {Smareglia}, {Smart}, {Smith}, {Solano}, {Solitro},
  {Sordo}, {Soria Nieto}, {Souchay}, {Spagna}, {Spoto}, {Stampa}, {Steele},
  {Steidelm{\"u}ller}, {Stephenson}, {Stoev}, {Suess}, {S{\"u}veges}, {Surdej},
  {Szabados}, {Szegedi-Elek}, {Tapiador}, {Taris}, {Tauran}, {Taylor},
  {Teixeira}, {Terrett}, {Tingley}, {Trager}, {Turon}, {Ulla}, {Utrilla},
  {Valentini}, {van Elteren}, {Van Hemelryck}, {van Leeuwen}, {Varadi},
  {Vecchiato}, {Veljanoski}, {Via}, {Vicente}, {Vogt}, {Voss}, {Votruba},
  {Voutsinas}, {Walmsley}, {Weiler}, {Weingrill}, {Werner}, {Wevers},
  {Whitehead}, {Wyrzykowski}, {Yoldas}, {{\v{Z}}erjal}, {Zucker}, {Zurbach},
  {Zwitter}, {Alecu}, {Allen}, {Allende Prieto}, {Amorim},
  {Anglada-Escud{\'e}}, {Arsenijevic}, {Azaz}, {Balm}, {Beck}, {Bernstein},
  {Bigot}, {Bijaoui}, {Blasco}, {Bonfigli}, {Bono}, {Boudreault}, {Bressan},
  {Brown}, {Brunet}, {Bunclark}, {Buonanno}, {Butkevich}, {Carret}, {Carrion},
  {Chemin}, {Ch{\'e}reau}, {Corcione}, {Darmigny}, {de Boer}, {de Teodoro}, {de
  Zeeuw}, {Delle Luche}, {Domingues}, {Dubath}, {Fodor}, {Fr{\'e}zouls},
  {Fries}, {Fustes}, {Fyfe}, {Gallardo}, {Gallegos}, {Gardiol}, {Gebran},
  {Gomboc}, {G{\'o}mez}, {Grux}, {Gueguen}, {Heyrovsky}, {Hoar}, {Iannicola},
  {Isasi Parache}, {Janotto}, {Joliet}, {Jonckheere}, {Keil}, {Kim},
  {Klagyivik}, {Klar}, {Knude}, {Kochukhov}, {Kolka}, {Kos}, {Kutka}, {Lainey},
  {LeBouquin}, {Liu}, {Loreggia}, {Makarov}, {Marseille}, {Martayan},
  {Martinez-Rubi}, {Massart}, {Meynadier}, {Mignot}, {Munari}, {Nguyen},
  {Nordlander}, {Ocvirk}, {O'Flaherty}, {Olias Sanz}, {Ortiz}, {Osorio},
  {Oszkiewicz}, {Ouzounis}, {Palmer}, {Park}, {Pasquato}, {Peltzer}, {Peralta},
  {P{\'e}turaud}, {Pieniluoma}, {Pigozzi}, {Poels}, {Prat}, {Prod'homme},
  {Raison}, {Rebordao}, {Risquez}, {Rocca-Volmerange}, {Rosen}, {Ruiz-Fuertes},
  {Russo}, {Sembay}, {Serraller Vizcaino}, {Short}, {Siebert}, {Silva},
  {Sinachopoulos}, {Slezak}, {Soffel}, {Sosnowska}, {Strai{\v{z}}ys}, {ter
  Linden}, {Terrell}, {Theil}, {Tiede}, {Troisi}, {Tsalmantza}, {Tur},
  {Vaccari}, {Vachier}, {Valles}, {Van Hamme}, {Veltz}, {Virtanen}, {Wallut},
  {Wichmann}, {Wilkinson}, {Ziaeepour}, \& {Zschocke}}]{Gaia2016}
{Gaia Collaboration}, {Prusti}, T., {de Bruijne}, J.~H.~J., {et~al.} 2016,
  \aap, 595, A1, \dodoi{10.1051/0004-6361/201629272}

\bibitem[{{Gaia Collaboration} {et~al.}(2018){Gaia Collaboration}, {Brown},
  {Vallenari}, {Prusti}, {de Bruijne}, {Babusiaux}, {Bailer-Jones}, {Biermann},
  {Evans}, {Eyer}, {Jansen}, {Jordi}, {Klioner}, {Lammers}, {Lindegren},
  {Luri}, {Mignard}, {Panem}, {Pourbaix}, {Randich}, {Sartoretti}, {Siddiqui},
  {Soubiran}, {van Leeuwen}, {Walton}, {Arenou}, {Bastian}, {Cropper},
  {Drimmel}, {Katz}, {Lattanzi}, {Bakker}, {Cacciari}, {Casta{\~n}eda},
  {Chaoul}, {Cheek}, {De Angeli}, {Fabricius}, {Guerra}, {Holl}, {Masana},
  {Messineo}, {Mowlavi}, {Nienartowicz}, {Panuzzo}, {Portell}, {Riello},
  {Seabroke}, {Tanga}, {Th{\'e}venin}, {Gracia-Abril}, {Comoretto},
  {Garcia-Reinaldos}, {Teyssier}, {Altmann}, {Andrae}, {Audard},
  {Bellas-Velidis}, {Benson}, {Berthier}, {Blomme}, {Burgess}, {Busso},
  {Carry}, {Cellino}, {Clementini}, {Clotet}, {Creevey}, {Davidson}, {De
  Ridder}, {Delchambre}, {Dell'Oro}, {Ducourant},
  {Fern{\'a}ndez-Hern{\'a}ndez}, {Fouesneau}, {Fr{\'e}mat}, {Galluccio},
  {Garc{\'\i}a-Torres}, {Gonz{\'a}lez-N{\'u}{\~n}ez}, {Gonz{\'a}lez-Vidal},
  {Gosset}, {Guy}, {Halbwachs}, {Hambly}, {Harrison}, {Hern{\'a}ndez},
  {Hestroffer}, {Hodgkin}, {Hutton}, {Jasniewicz}, {Jean-Antoine-Piccolo},
  {Jordan}, {Korn}, {Krone-Martins}, {Lanzafame}, {Lebzelter}, {L{\"o}ffler},
  {Manteiga}, {Marrese}, {Mart{\'\i}n-Fleitas}, {Moitinho}, {Mora}, {Muinonen},
  {Osinde}, {Pancino}, {Pauwels}, {Petit}, {Recio-Blanco}, {Richards},
  {Rimoldini}, {Robin}, {Sarro}, {Siopis}, {Smith}, {Sozzetti}, {S{\"u}veges},
  {Torra}, {van Reeven}, {Abbas}, {Abreu Aramburu}, {Accart}, {Aerts},
  {Altavilla}, {{\'A}lvarez}, {Alvarez}, {Alves}, {Anderson}, {Andrei},
  {Anglada Varela}, {Antiche}, {Antoja}, {Arcay}, {Astraatmadja}, {Bach},
  {Baker}, {Balaguer-N{\'u}{\~n}ez}, {Balm}, {Barache}, {Barata}, {Barbato},
  {Barblan}, {Barklem}, {Barrado}, {Barros}, {Barstow}, {Bartholom{\'e}
  Mu{\~n}oz}, {Bassilana}, {Becciani}, {Bellazzini}, {Berihuete}, {Bertone},
  {Bianchi}, {Bienaym{\'e}}, {Blanco-Cuaresma}, {Boch}, {Boeche}, {Bombrun},
  {Borrachero}, {Bossini}, {Bouquillon}, {Bourda}, {Bragaglia}, {Bramante},
  {Breddels}, {Bressan}, {Brouillet}, {Br{\"u}semeister}, {Brugaletta},
  {Bucciarelli}, {Burlacu}, {Busonero}, {Butkevich}, {Buzzi}, {Caffau},
  {Cancelliere}, {Cannizzaro}, {Cantat-Gaudin}, {Carballo}, {Carlucci},
  {Carrasco}, {Casamiquela}, {Castellani}, {Castro-Ginard}, {Charlot},
  {Chemin}, {Chiavassa}, {Cocozza}, {Costigan}, {Cowell}, {Crifo}, {Crosta},
  {Crowley}, {Cuypers}, {Dafonte}, {Damerdji}, {Dapergolas}, {David}, {David},
  {de Laverny}, {De Luise}, {De March}, {de Martino}, {de Souza}, {de Torres},
  {Debosscher}, {del Pozo}, {Delbo}, {Delgado}, {Delgado}, {Di Matteo},
  {Diakite}, {Diener}, {Distefano}, {Dolding}, {Drazinos}, {Dur{\'a}n},
  {Edvardsson}, {Enke}, {Eriksson}, {Esquej}, {Eynard Bontemps}, {Fabre},
  {Fabrizio}, {Faigler}, {Falc{\~a}o}, {Farr{\`a}s Casas}, {Federici},
  {Fedorets}, {Fernique}, {Figueras}, {Filippi}, {Findeisen}, {Fonti},
  {Fraile}, {Fraser}, {Fr{\'e}zouls}, {Gai}, {Galleti}, {Garabato},
  {Garc{\'\i}a-Sedano}, {Garofalo}, {Garralda}, {Gavel}, {Gavras}, {Gerssen},
  {Geyer}, {Giacobbe}, {Gilmore}, {Girona}, {Giuffrida}, {Glass}, {Gomes},
  {Granvik}, {Gueguen}, {Guerrier}, {Guiraud}, {Guti{\'e}rrez-S{\'a}nchez},
  {Haigron}, {Hatzidimitriou}, {Hauser}, {Haywood}, {Heiter}, {Helmi}, {Heu},
  {Hilger}, {Hobbs}, {Hofmann}, {Holland}, {Huckle}, {Hypki}, {Icardi},
  {Jan{\ss}en}, {Jevardat de Fombelle}, {Jonker}, {Juh{\'a}sz}, {Julbe},
  {Karampelas}, {Kewley}, {Klar}, {Kochoska}, {Kohley}, {Kolenberg},
  {Kontizas}, {Kontizas}, {Koposov}, {Kordopatis}, {Kostrzewa-Rutkowska},
  {Koubsky}, {Lambert}, {Lanza}, {Lasne}, {Lavigne}, {Le Fustec}, {Le
  Poncin-Lafitte}, {Lebreton}, {Leccia}, {Leclerc}, {Lecoeur-Taibi},
  {Lenhardt}, {Leroux}, {Liao}, {Licata}, {Lindstr{\o}m}, {Lister}, {Livanou},
  {Lobel}, {L{\'o}pez}, {Managau}, {Mann}, {Mantelet}, {Marchal}, {Marchant},
  {Marconi}, {Marinoni}, {Marschalk{\'o}}, {Marshall}, {Martino}, {Marton},
  {Mary}, {Massari}, {Matijevi{\v{c}}}, {Mazeh}, {McMillan}, {Messina},
  {Michalik}, {Millar}, {Molina}, {Molinaro}, {Moln{\'a}r}, {Montegriffo},
  {Mor}, {Morbidelli}, {Morel}, {Morris}, {Mulone}, {Muraveva}, {Musella},
  {Nelemans}, {Nicastro}, {Noval}, {O'Mullane}, {Ord{\'e}novic},
  {Ord{\'o}{\~n}ez-Blanco}, {Osborne}, {Pagani}, {Pagano}, {Pailler},
  {Palacin}, {Palaversa}, {Panahi}, {Pawlak}, {Piersimoni}, {Pineau}, {Plachy},
  {Plum}, {Poggio}, {Poujoulet}, {Pr{\v{s}}a}, {Pulone}, {Racero}, {Ragaini},
  {Rambaux}, {Ramos-Lerate}, {Regibo}, {Reyl{\'e}}, {Riclet}, {Ripepi}, {Riva},
  {Rivard}, {Rixon}, {Roegiers}, {Roelens}, {Romero-G{\'o}mez}, {Rowell},
  {Royer}, {Ruiz-Dern}, {Sadowski}, {Sagrist{\`a} Sell{\'e}s}, {Sahlmann},
  {Salgado}, {Salguero}, {Sanna}, {Santana-Ros}, {Sarasso}, {Savietto},
  {Schultheis}, {Sciacca}, {Segol}, {Segovia}, {S{\'e}gransan}, {Shih},
  {Siltala}, {Silva}, {Smart}, {Smith}, {Solano}, {Solitro}, {Sordo}, {Soria
  Nieto}, {Souchay}, {Spagna}, {Spoto}, {Stampa}, {Steele},
  {Steidelm{\"u}ller}, {Stephenson}, {Stoev}, {Suess}, {Surdej}, {Szabados},
  {Szegedi-Elek}, {Tapiador}, {Taris}, {Tauran}, {Taylor}, {Teixeira},
  {Terrett}, {Teyssandier}, {Thuillot}, {Titarenko}, {Torra Clotet}, {Turon},
  {Ulla}, {Utrilla}, {Uzzi}, {Vaillant}, {Valentini}, {Valette}, {van Elteren},
  {Van Hemelryck}, {van Leeuwen}, {Vaschetto}, {Vecchiato}, {Veljanoski},
  {Viala}, {Vicente}, {Vogt}, {von Essen}, {Voss}, {Votruba}, {Voutsinas},
  {Walmsley}, {Weiler}, {Wertz}, {Wevers}, {Wyrzykowski}, {Yoldas},
  {{\v{Z}}erjal}, {Ziaeepour}, {Zorec}, {Zschocke}, {Zucker}, {Zurbach}, \&
  {Zwitter}}]{GaiaDR2}
{Gaia Collaboration}, {Brown}, A.~G.~A., {Vallenari}, A., {et~al.} 2018, \aap,
  616, A1, \dodoi{10.1051/0004-6361/201833051}

\bibitem[{{Garc{\'\i}a Mu{\~n}oz} {et~al.}(2020){Garc{\'\i}a Mu{\~n}oz},
  {Youngblood}, {Fossati}, {Gandolfi}, {Cabrera}, \& {Rauer}}]{GarciaMunoz2020}
{Garc{\'\i}a Mu{\~n}oz}, A., {Youngblood}, A., {Fossati}, L., {et~al.} 2020,
  \apjl, 888, L21, \dodoi{10.3847/2041-8213/ab61ff}

\bibitem[{{Ghezzi} {et~al.}(2010){Ghezzi}, {Cunha}, {Smith}, {de Ara{\'u}jo},
  {Schuler}, \& {de la Reza}}]{Ghezzi2010}
{Ghezzi}, L., {Cunha}, K., {Smith}, V.~V., {et~al.} 2010, \apj, 720, 1290,
  \dodoi{10.1088/0004-637X/720/2/1290}

\bibitem[{Guinan {et~al.}(2016)Guinan, Engle, Durbin, et~al
  {Anglada-Escud{\'{e}} G., Arriagada P.}, et~al {Anglada-Escud{\'{e}} G.,
  Arriagada P.}, et~al {Anglada-Escud{\'{e}} G., Tuomi M.}, et~al
  {Astudillo-Defru N., Bonfils X.}, R., R., {Barrado y Navascu{\'{e}}s D.,
  Stauffer J. R.}, J.-P., C., K., et~al {Bochanski J. J., Hawley S. L.}, et~al
  {Bonfils X., Delfosse X.}, {Buccino A. P.}, D., et~al {Campante T. L.,
  Barclay T.}, et~al {Chaplin W. J., Lund M. N.}, N., D., S., Evren, et~al
  {Deshpande R., Blake C. H.}, D., N., D., {Doyle J. G.}, B., D., Charbonneau,
  D., Charbonneau, R., Barnes, {Durbin A. J.}, G., J., J., et~Al, ed~Qain~S.,
  {Engle S. G.}, G., ESA, et~al {Forveille T., Bonfils X.}, et~al {France K.,
  Froning C. S.}, et~al {France K., Shkolnik E.}, et~al {Fuhrmeister B.,
  Lalitha S.}, {Garc{\'{e}}s A.}, I., E., {Gomes da Silva J., Santos N. C.,
  Boisse I.}, C., et~al {Grie{\ss}meier J.-M., Stadelmann A.}, {G{\"{u}}del M.
  ed Kamp I.}, D., {Guinan E. F.}, G., {Guinan E. F.}, M., et~al {Haisch B. M.,
  Linsky J. L.}, et~al {Henry T. J., Jao W.-C.}, {Hilton E. J., West A. A.},
  F., {Hunt-Walker N. M., Hilton E. J., Kowalski A. F.}, M., {Irwin J. M.,
  Berta-Thompson Z. K.}, C., M., G{\"{u}}del, S., C., K., Stepien, et~al
  {Klutsch A., Freire Ferrero R.}, {Kotoneva E., Innanen K., Dawson P. C.}, M.,
  et~al {Kowalski A. F., Hawley S. L.}, {Lammer H., Terada N.}, K., {Linsky J.
  L.}, K., et~al {Maehara H., Shibayama T.}, M., P., A., L., {Micela G.}, S.,
  {Mullan D. J.}, J., et~al {Navarrete C., Chanam{\'{e}} J.}, {Nidever D. L.,
  Marcy G. W., Butler R. P.}, S., et~al {Notsu Y., Shibayama T.}, et~al
  {Nysewander M., Reichart D. E.}, S., T., et~al {Osten R. A., Godet O.},
  {Osten R. A., Hawley S. L., Allred J. C.}, C., F., et~al {Pepper J., Stassun
  K.}, et~al {Reichart D., Nysewander M.}, G., Basri, S., Mikkola, {Ribas I.,
  Guinan E. F.}, M., {Robertson P.}, S., {Rugheimer S., Kaltenegger L., Segura
  A.}, S., C., Liefke, A., {S{\'{e}}gransan D., Kervella P.}, D., {Segura A.,
  Walkowicz L. M., Meadows V.}, S., et~al {Silva Aguirre V., Davies G. R.}, F.,
  {Tuomi M., Jones H. R. A., Barnes J. R.}, S., F., et~al {Vaughan A. H.,
  Preston G. W.}, R., et~al {West A. A., Hawley S. L.}, et~al {West A. A.,
  Weisenburger K. L.}, C., {Wood B. E., Linsky J. L.}, P., {Wood B. E.,
  Redfield S., Linsky J. L.}, P., G., Wallerstein, {Wylie-de Boer E.}, M.,
  {Youngblood A.}, O., {Zhao J. K., Oswalt T. D., Rudkin M.}, \&
  Q.}]{Guinan2016}
Guinan, E.~F., Engle, S.~G., Durbin, A., {et~al.} 2016, The Astrophysical
  Journal, 821, 81.
\newblock
  \url{http://stacks.iop.org/0004-637X/821/i=2/a=81?key=crossref.94bd7f8e4f585d83b9f30bfbe7430bb9}

\bibitem[{{Gun{\'a}r} {et~al.}(2021){Gun{\'a}r}, {Koza}, {Schwartz}, {Heinzel},
  \& {Liu}}]{Gunar2021}
{Gun{\'a}r}, S., {Koza}, J., {Schwartz}, P., {Heinzel}, P., \& {Liu}, W. 2021,
  \apjs, 255, 16, \dodoi{10.3847/1538-4365/ac07ab}

\bibitem[{{Gun{\'a}r} {et~al.}(2020){Gun{\'a}r}, {Schwartz}, {Koza}, \&
  {Heinzel}}]{Gunar2020}
{Gun{\'a}r}, S., {Schwartz}, P., {Koza}, J., \& {Heinzel}, P. 2020, \aap, 644,
  A109, \dodoi{10.1051/0004-6361/202039348}

\bibitem[{{Howard} \& {Fulton}(2016)}]{Howard2016}
{Howard}, A.~W., \& {Fulton}, B.~J. 2016, \pasp, 128, 114401,
  \dodoi{10.1088/1538-3873/128/969/114401}

\bibitem[{Hunter(2007)}]{Hunter2007}
Hunter, J.~D. 2007, Computing in Science {\&} Engineering, 9, 90,
  \dodoi{10.1109/MCSE.2007.55}

\bibitem[{{Judge} {et~al.}(2020){Judge}, {Kleint}, {Leenaarts}, {Sukhorukov},
  \& {Vial}}]{Judge2020}
{Judge}, P.~G., {Kleint}, L., {Leenaarts}, J., {Sukhorukov}, A.~V., \& {Vial},
  J.-C. 2020, \apj, 901, 32, \dodoi{10.3847/1538-4357/abadf4}

\bibitem[{{Kesseli} {et~al.}(2019){Kesseli}, {Kirkpatrick}, {Fajardo-Acosta},
  {Penny}, {Gaudi}, {Veyette}, {Boeshaar}, {Henderson}, {Cushing},
  {Calchi-Novati}, {Shvartzvald}, \& {Muirhead}}]{Kesseli2019}
{Kesseli}, A.~Y., {Kirkpatrick}, J.~D., {Fajardo-Acosta}, S.~B., {et~al.} 2019,
  \aj, 157, 63, \dodoi{10.3847/1538-3881/aae982}

\bibitem[{{Kotoneva} {et~al.}(2005){Kotoneva}, {Innanen}, {Dawson}, {Wood}, \&
  {De Robertis}}]{Kotoneva2005}
{Kotoneva}, E., {Innanen}, K., {Dawson}, P.~C., {Wood}, P.~R., \& {De
  Robertis}, M.~M. 2005, \aap, 438, 957, \dodoi{10.1051/0004-6361:20042287}

\bibitem[{{Linsky} {et~al.}(1979){Linsky}, {Worden}, {McClintock}, \&
  {Robertson}}]{Linsky1979a}
{Linsky}, J.~L., {Worden}, S.~P., {McClintock}, W., \& {Robertson}, R.~M. 1979,
  \apjs, 41, 47, \dodoi{10.1086/190607}

\bibitem[{{Linsky} {et~al.}(2006){Linsky}, {Draine}, {Moos}, {Jenkins}, {Wood},
  {Oliveira}, {Blair}, {Friedman}, {Gry}, {Knauth}, {Kruk}, {Lacour}, {Lehner},
  {Redfield}, {Shull}, {Sonneborn}, \& {Williger}}]{Linsky2006}
{Linsky}, J.~L., {Draine}, B.~T., {Moos}, H.~W., {et~al.} 2006, \apj, 647,
  1106, \dodoi{10.1086/505556}

\bibitem[{{Linsky} {et~al.}(2020){Linsky}, {Wood}, {Youngblood}, {Brown},
  {Froning}, {France}, {Buccino}, {Cranmer}, {Mauas}, {Miguel}, {Pineda},
  {Rugheimer}, {Vieytes}, {Wheatley}, \& {Wilson}}]{Linsky2020}
{Linsky}, J.~L., {Wood}, B.~E., {Youngblood}, A., {et~al.} 2020, \apj, 902, 3,
  \dodoi{10.3847/1538-4357/abb36f}

\bibitem[{{Mann} {et~al.}(2019){Mann}, {Dupuy}, {Kraus}, {Gaidos}, {Ansdell},
  {Ireland}, {Rizzuto}, {Hung}, {Dittmann}, {Factor}, {Feiden}, {Martinez},
  {Ru{\'\i}z-Rodr{\'\i}guez}, \& {Thao}}]{Mann2019}
{Mann}, A.~W., {Dupuy}, T., {Kraus}, A.~L., {et~al.} 2019, \apj, 871, 63,
  \dodoi{10.3847/1538-4357/aaf3bc}

\bibitem[{{Mar{\'\i}n-Franch} {et~al.}(2009){Mar{\'\i}n-Franch}, {Aparicio},
  {Piotto}, {Rosenberg}, {Chaboyer}, {Sarajedini}, {Siegel}, {Anderson},
  {Bedin}, {Dotter}, {Hempel}, {King}, {Majewski}, {Milone}, {Paust}, \&
  {Reid}}]{MarinFranch2009}
{Mar{\'\i}n-Franch}, A., {Aparicio}, A., {Piotto}, G., {et~al.} 2009, \apj,
  694, 1498, \dodoi{10.1088/0004-637X/694/2/1498}

\bibitem[{McLean {et~al.}(1994)McLean, Mitchell, \& Swanston}]{McLean1994}
McLean, A.~B., Mitchell, C. E.~J., \& Swanston, D.~M. 1994, Journal of Electron
  Spectroscopy and Related Phenomena, 69, 125,
  \dodoi{10.1016/0368-2048(94)02189-7}

\bibitem[{Miguel {et~al.}(2015)Miguel, Kaltenegger, Linsky, \&
  Rugheimer}]{Miguel2015}
Miguel, Y., Kaltenegger, L., Linsky, J.~L., \& Rugheimer, S. 2015, Monthly
  Notices of the Royal Astronomical Society, 446, 345,
  \dodoi{10.1093/mnras/stu2107}

\bibitem[{{Miller-Ricci Kempton} {et~al.}(2012){Miller-Ricci Kempton},
  {Zahnle}, \& {Fortney}}]{MillerRicciKempton2012}
{Miller-Ricci Kempton}, E., {Zahnle}, K., \& {Fortney}, J.~J. 2012, \apj, 745,
  3, \dodoi{10.1088/0004-637X/745/1/3}

\bibitem[{{Neves} {et~al.}(2013){Neves}, {Bonfils}, {Santos}, {Delfosse},
  {Forveille}, {Allard}, \& {Udry}}]{Neves2013}
{Neves}, V., {Bonfils}, X., {Santos}, N.~C., {et~al.} 2013, \aap, 551, A36,
  \dodoi{10.1051/0004-6361/201220574}

\bibitem[{{Nidever} {et~al.}(2002){Nidever}, {Marcy}, {Butler}, {Fischer}, \&
  {Vogt}}]{Nidever2002}
{Nidever}, D.~L., {Marcy}, G.~W., {Butler}, R.~P., {Fischer}, D.~A., \& {Vogt},
  S.~S. 2002, \apjs, 141, 503, \dodoi{10.1086/340570}

\bibitem[{{Peacock} {et~al.}(2019{\natexlab{a}}){Peacock}, {Barman},
  {Shkolnik}, {Hauschildt}, \& {Baron}}]{Peacock2019_trappist}
{Peacock}, S., {Barman}, T., {Shkolnik}, E.~L., {Hauschildt}, P.~H., \&
  {Baron}, E. 2019{\natexlab{a}}, \apj, 871, 235,
  \dodoi{10.3847/1538-4357/aaf891}

\bibitem[{{Peacock} {et~al.}(2019{\natexlab{b}}){Peacock}, {Barman},
  {Shkolnik}, {Hauschildt}, {Baron}, \& {Fuhrmeister}}]{Peacock_2019b}
{Peacock}, S., {Barman}, T., {Shkolnik}, E.~L., {et~al.} 2019{\natexlab{b}},
  \apj, 886, 77, \dodoi{10.3847/1538-4357/ab4f6f}

\bibitem[{{Pecaut} \& {Mamajek}(2013)}]{Pecaut2013}
{Pecaut}, M.~J., \& {Mamajek}, E.~E. 2013, \apjs, 208, 9,
  \dodoi{10.1088/0067-0049/208/1/9}

\bibitem[{{Pepe} {et~al.}(2011){Pepe}, {Lovis}, {S{\'e}gransan}, {Benz},
  {Bouchy}, {Dumusque}, {Mayor}, {Queloz}, {Santos}, \& {Udry}}]{Pepe2011}
{Pepe}, F., {Lovis}, C., {S{\'e}gransan}, D., {et~al.} 2011, \aap, 534, A58,
  \dodoi{10.1051/0004-6361/201117055}

\bibitem[{Perez \& Granger(2007)}]{Perez2007}
Perez, F., \& Granger, B.~E. 2007, Computing in Science {\&} Engineering, 9,
  21, \dodoi{10.1109/MCSE.2007.53}

\bibitem[{{Pineda} {et~al.}(2021){Pineda}, {Youngblood}, \&
  {France}}]{Pineda2021_MUSS}
{Pineda}, J.~S., {Youngblood}, A., \& {France}, K. 2021, arXiv e-prints,
  arXiv:2106.07656.
\newblock \doarXiv{2106.07656}

\bibitem[{{Proffitt} {et~al.}(2017){Proffitt}, {Monroe}, \&
  {Dressel}}]{Proffitt2017}
{Proffitt}, C.~R., {Monroe}, T., \& {Dressel}, L. 2017, {Status of the STIS
  Instrument Focus}, Space Telescope STIS Instrument Science Report

\bibitem[{Redfield \& Linsky(2008)}]{Redfield2008}
Redfield, S., \& Linsky, J.~L. 2008, The Astrophysical Journal, 673, 283,
  \dodoi{10.1086/524002}

\bibitem[{{Reiners} \& {Mohanty}(2012)}]{Reiners2012}
{Reiners}, A., \& {Mohanty}, S. 2012, \apj, 746, 43,
  \dodoi{10.1088/0004-637X/746/1/43}

\bibitem[{{Ribas} {et~al.}(2018){Ribas}, {Tuomi}, {Reiners}, {Butler},
  {Morales}, {Perger}, {Dreizler}, {Rodr{\'\i}guez-L{\'o}pez}, {Gonz{\'a}lez
  Hern{\'a}ndez}, {Rosich}, {Feng}, {Trifonov}, {Vogt}, {Caballero}, {Hatzes},
  {Herrero}, {Jeffers}, {Lafarga}, {Murgas}, {Nelson}, {Rodr{\'\i}guez},
  {Strachan}, {Tal-Or}, {Teske}, {Toledo-Padr{\'o}n}, {Zechmeister},
  {Quirrenbach}, {Amado}, {Azzaro}, {B{\'e}jar}, {Barnes}, {Berdi{\~n}as},
  {Burt}, {Coleman}, {Cort{\'e}s-Contreras}, {Crane}, {Engle}, {Guinan},
  {Haswell}, {Henning}, {Holden}, {Jenkins}, {Jones}, {Kaminski}, {Kiraga},
  {K{\"u}rster}, {Lee}, {L{\'o}pez-Gonz{\'a}lez}, {Montes}, {Morin}, {Ofir},
  {Pall{\'e}}, {Rebolo}, {Reffert}, {Schweitzer}, {Seifert}, {Shectman},
  {Staab}, {Street}, {Su{\'a}rez Mascare{\~n}o}, {Tsapras}, {Wang}, \&
  {Anglada-Escud{\'e}}}]{Ribas2018}
{Ribas}, I., {Tuomi}, M., {Reiners}, A., {et~al.} 2018, \nat, 563, 365,
  \dodoi{10.1038/s41586-018-0677-y}

\bibitem[{{Riley} {et~al.}(2018){Riley}, {Monroe}, \& {Lockwood}}]{Riley2018}
{Riley}, A., {Monroe}, T., \& {Lockwood}, S. 2018, {Impacts of focus on aspects
  of STIS UV Spectroscopy}, Space Telescope STIS Instrument Science Report

\bibitem[{Robitaille {et~al.}(2013)Robitaille, Tollerud, Greenfield,
  Droettboom, Bray, Aldcroft, Davis, Ginsburg, Price-Whelan, Kerzendorf,
  Conley, Crighton, Barbary, Muna, Ferguson, Grollier, Parikh, Nair,
  G{\"{u}}nther, Deil, Woillez, Conseil, Kramer, Turner, Singer, Fox, Weaver,
  Zabalza, Edwards, {Azalee Bostroem}, Burke, Casey, Crawford, Dencheva, Ely,
  Jenness, Labrie, Lim, Pierfederici, Pontzen, Ptak, Refsdal, Servillat, \&
  Streicher}]{Robitaille2013}
Robitaille, T.~P., Tollerud, E.~J., Greenfield, P., {et~al.} 2013, Astronomy
  {\&} Astrophysics, 558, A33, \dodoi{10.1051/0004-6361/201322068}

\bibitem[{{Rojas-Ayala} {et~al.}(2012){Rojas-Ayala}, {Covey}, {Muirhead}, \&
  {Lloyd}}]{RojasAyala2012}
{Rojas-Ayala}, B., {Covey}, K.~R., {Muirhead}, P.~S., \& {Lloyd}, J.~P. 2012,
  \apj, 748, 93, \dodoi{10.1088/0004-637X/748/2/93}

\bibitem[{{Rosenthal} {et~al.}(2021){Rosenthal}, {Fulton}, {Hirsch},
  {Isaacson}, {Howard}, {Dedrick}, {Sherstyuk}, {Blunt}, {Petigura}, {Knutson},
  {Behmard}, {Chontos}, {Crepp}, {Crossfield}, {Dalba}, {Fischer}, {Henry},
  {Kane}, {Kosiarek}, {Marcy}, {Rubenzahl}, {Weiss}, \&
  {Wright}}]{Rosenthal2021}
{Rosenthal}, L.~J., {Fulton}, B.~J., {Hirsch}, L.~A., {et~al.} 2021, \apjs,
  255, 8, \dodoi{10.3847/1538-4365/abe23c}

\bibitem[{{Rugheimer} {et~al.}(2015){Rugheimer}, {Kaltenegger}, {Segura},
  {Linsky}, \& {Mohanty}}]{Rugheimer2015}
{Rugheimer}, S., {Kaltenegger}, L., {Segura}, A., {Linsky}, J., \& {Mohanty},
  S. 2015, \apj, 809, 57, \dodoi{10.1088/0004-637X/809/1/57}

\bibitem[{{Schmit} {et~al.}(2015){Schmit}, {Bryans}, {De Pontieu}, {McIntosh},
  {Leenaarts}, \& {Carlsson}}]{Schmit2015}
{Schmit}, D., {Bryans}, P., {De Pontieu}, B., {et~al.} 2015, \apj, 811, 127,
  \dodoi{10.1088/0004-637X/811/2/127}

\bibitem[{{Schneider} {et~al.}(2019){Schneider}, {Shkolnik}, {Barman}, \&
  {Loyd}}]{Schneider2019}
{Schneider}, A.~C., {Shkolnik}, E.~L., {Barman}, T.~S., \& {Loyd}, R.~P. 2019,
  \apj, 886, 19, \dodoi{10.3847/1538-4357/ab48de}

\bibitem[{{Short} \& {Doyle}(1998)}]{Short1998}
{Short}, C.~I., \& {Doyle}, J.~G. 1998, \aap, 336, 613

\bibitem[{{Soubiran} {et~al.}(2018){Soubiran}, {Jasniewicz}, {Chemin},
  {Zurbach}, {Brouillet}, {Panuzzo}, {Sartoretti}, {Katz}, {Le Campion},
  {Marchal}, {Hestroffer}, {Th{\'e}venin}, {Crifo}, {Udry}, {Cropper},
  {Seabroke}, {Viala}, {Benson}, {Blomme}, {Jean-Antoine}, {Huckle}, {Smith},
  {Baker}, {Damerdji}, {Dolding}, {Fr{\'e}mat}, {Gosset}, {Guerrier}, {Guy},
  {Haigron}, {Jan{\ss}en}, {Plum}, {Fabre}, {Lasne}, {Pailler}, {Panem},
  {Riclet}, {Royer}, {Tauran}, {Zwitter}, {Gueguen}, \& {Turon}}]{Soubiran2018}
{Soubiran}, C., {Jasniewicz}, G., {Chemin}, L., {et~al.} 2018, \aap, 616, A7,
  \dodoi{10.1051/0004-6361/201832795}

\bibitem[{{Tian} {et~al.}(2009{\natexlab{a}}){Tian}, {Curdt}, {Marsch}, \&
  {Sch{\"u}hle}}]{Tian2009_LyALyB}
{Tian}, H., {Curdt}, W., {Marsch}, E., \& {Sch{\"u}hle}, U. 2009{\natexlab{a}},
  \aap, 504, 239, \dodoi{10.1051/0004-6361/200811445}

\bibitem[{{Tian} {et~al.}(2009{\natexlab{b}}){Tian}, {Curdt}, {Teriaca},
  {Landi}, \& {Marsch}}]{Tian2009_transitionregion_sunspots}
{Tian}, H., {Curdt}, W., {Teriaca}, L., {Landi}, E., \& {Marsch}, E.
  2009{\natexlab{b}}, \aap, 505, 307, \dodoi{10.1051/0004-6361/200912114}

\bibitem[{{Tian} {et~al.}(2009{\natexlab{c}}){Tian}, {Teriaca}, {Curdt}, \&
  {Vial}}]{Tian2009_coronalholes}
{Tian}, H., {Teriaca}, L., {Curdt}, W., \& {Vial}, J.-C. 2009{\natexlab{c}},
  \apjl, 703, L152, \dodoi{10.1088/0004-637X/703/2/L152}

\bibitem[{{Tilipman} {et~al.}(2021){Tilipman}, {Vieytes}, {Linsky}, {Buccino},
  \& {France}}]{Tilipman2021}
{Tilipman}, D., {Vieytes}, M., {Linsky}, J.~L., {Buccino}, A.~P., \& {France},
  K. 2021, \apj, 909, 61, \dodoi{10.3847/1538-4357/abd62f}

\bibitem[{van~der Walt {et~al.}(2011)van~der Walt, Colbert, \&
  Varoquaux}]{VanderWalt2011}
van~der Walt, S., Colbert, S.~C., \& Varoquaux, G. 2011, Computing in Science
  {\&} Engineering, 13, 22, \dodoi{10.1109/MCSE.2011.37}

\bibitem[{{Vernazza} {et~al.}(1981){Vernazza}, {Avrett}, \&
  {Loeser}}]{Vernazza1981}
{Vernazza}, J.~E., {Avrett}, E.~H., \& {Loeser}, R. 1981, \apjs, 45, 635,
  \dodoi{10.1086/190731}

\bibitem[{{Wilhelm} {et~al.}(1997){Wilhelm}, {Lemaire}, {Curdt}, {Schuhle},
  {Marsch}, {Poland}, {Jordan}, {Thomas}, {Hassler}, {Huber}, {Vial}, {Kuhne},
  {Siegmund}, {Gabriel}, {Timothy}, {Grewing}, {Feldman}, {Hollandt}, \&
  {Brekke}}]{Wilhelm1997}
{Wilhelm}, K., {Lemaire}, P., {Curdt}, W., {et~al.} 1997, \solphys, 170, 75,
  \dodoi{10.1023/A:1004923511980}

\bibitem[{Wilson \& Bappu(1957)}]{Wilson1957}
Wilson, O.~C., \& Bappu, M. K.~V. 1957, The Astrophysical Journal, 125, 661,
  \dodoi{10.1086/146339}

\bibitem[{{Wood} {et~al.}(2005){Wood}, {Redfield}, {Linsky}, {M{\"u}ller}, \&
  {Zank}}]{Wood2005}
{Wood}, B.~E., {Redfield}, S., {Linsky}, J.~L., {M{\"u}ller}, H.-R., \& {Zank},
  G.~P. 2005, \apjs, 159, 118, \dodoi{10.1086/430523}

\bibitem[{{Wood} {et~al.}(2021){Wood}, {M{\"u}ller}, {Redfield}, {Konow},
  {Vannier}, {Linsky}, {Youngblood}, {Vidotto}, {Jardine},
  {Alvarado-G{\'o}mez}, \& {Drake}}]{Wood2021}
{Wood}, B.~E., {M{\"u}ller}, H.-R., {Redfield}, S., {et~al.} 2021, \apj, 915,
  37, \dodoi{10.3847/1538-4357/abfda5}

\bibitem[{{Youngblood} {et~al.}(2021){Youngblood}, {Pineda}, \&
  {France}}]{Youngblood2021}
{Youngblood}, A., {Pineda}, J.~S., \& {France}, K. 2021, \apj, 911, 112,
  \dodoi{10.3847/1538-4357/abe8d8}

\bibitem[{{Youngblood} {et~al.}(2016){Youngblood}, {France}, {Loyd}, {Linsky},
  {Redfield}, {Schneider}, {Wood}, {Brown}, {Froning}, {Miguel}, {Rugheimer},
  \& {Walkowicz}}]{Youngblood2016}
{Youngblood}, A., {France}, K., {Loyd}, R.~O.~P., {et~al.} 2016, \apj, 824,
  101, \dodoi{10.3847/0004-637X/824/2/101}

\bibitem[{{Zacharias} {et~al.}(2013){Zacharias}, {Finch}, {Girard}, {Henden},
  {Bartlett}, {Monet}, \& {Zacharias}}]{Zacharias2013}
{Zacharias}, N., {Finch}, C.~T., {Girard}, T.~M., {et~al.} 2013, \aj, 145, 44,
  \dodoi{10.1088/0004-6256/145/2/44}

\bibitem[{{Zhang} {et~al.}(2021){Zhang}, {Knutson}, {Wang}, {Dai}, {dos
  Santos}, {Fossati}, {Henry}, {Ehrenreich}, {Alibert}, {Hoyer}, {Wilson}, \&
  {Bonfanti}}]{Zhang2021}
{Zhang}, M., {Knutson}, H.~A., {Wang}, L., {et~al.} 2021, arXiv e-prints,
  arXiv:2106.05273.
\newblock \doarXiv{2106.05273}

\end{thebibliography}
\bibliographystyle{aasjournal}

\end{document}